\documentclass[aps,prd,nofootinbib,superscriptaddress,preprintnumbers,longbibliography,eqsecnum,notitlepage,english]{revtex4-1}
\usepackage[export]{adjustbox}
\usepackage{amsfonts,amsmath,amssymb}
\usepackage{enumitem,mathtools}
\usepackage{array,multirow,booktabs,enumitem}
\usepackage{bbold,bbm,bm,nicefrac,dsfont}
\usepackage[dvipsnames,usenames]{color}
\usepackage{graphicx}
\usepackage{lipsum,float}
\usepackage[normalem]{ulem}
\usepackage[dvipsnames]{xcolor}
\usepackage[utf8]{inputenc}
\usepackage{hyperref}
\hypersetup{
    colorlinks = true,
    linkcolor = blue,
    urlcolor  = blue,
    citecolor = blue,
    anchorcolor = blue,
    pdftitle={Effects of Final State Interactions on Landau Singularities},
    pdfauthor={A. S. Sakthivasan, M. Mai, A. Rusetsky, and M. Döring}
}
\usepackage{cleveref} 
    \crefformat{figure}{Fig.~#2#1#3}
    \crefformat{table}{Tab.~#2#1#3}
    \crefformat{equation}{Eq.~(#2#1#3)}
    \crefformat{section}{Sect.~#2#1#3}

\renewcommand{\Re}{{\rm Re}}
\renewcommand{\Im}{{\rm Im}}
\newcommand{\MeV}{{\rm MeV}}
\newcommand{\GeV}{{\rm GeV}}
\graphicspath{{./plots/}}

\begin{document}
\title{Effects of Final State Interactions on Landau Singularities}

\author{A.~S.~Sakthivasan}
\email{sakthivasan@hiskp.uni-bonn.de}
\affiliation{Helmholtz-Institut f\"ur Strahlen- und Kernphysik (Theorie) and Bethe Center for Theoretical Physics, Universit\"at Bonn, 53115 Bonn, Germany}

\author{M.~Mai}
\email{mai@hiskp.uni-bonn.de}
\affiliation{Helmholtz-Institut f\"ur Strahlen- und Kernphysik (Theorie) and Bethe Center for Theoretical Physics, Universit\"at Bonn, 53115 Bonn, Germany}
\affiliation{Albert Einstein Center for Fundamental Physics, Institute for Theoretical Physics, University of Bern, Sidlerstrasse 5, 3012 Bern, Switzerland}
\affiliation{Institute for Nuclear Studies and Department of Physics, The George Washington University, Washington, DC 20052, USA}

\author{A.~Rusetsky}
\email{rusetsky@hiskp.uni-bonn.de}
\affiliation{Helmholtz-Institut f\"ur Strahlen- und Kernphysik (Theorie) and Bethe Center for Theoretical Physics, Universit\"at Bonn, 53115 Bonn, Germany}
\affiliation{Tbilisi State University, 0186 Tbilisi, Georgia}

\author{M.~D\"oring}
\email{doring@gwu.edu}
\affiliation{Institute for Nuclear Studies and Department of Physics, The George Washington University, Washington, DC 20052, USA}
\affiliation{Theory Center, Thomas Jefferson National Accelerator Facility, Newport News, VA 23606, USA}

\preprint{JLAB-THY-24-4209}


\begin{abstract}
In certain kinematic and particle mass configurations, triangle singularities may lead to line-shapes which mimic the effects of resonances. This well-known effect is scrutinized here in the presence of final-state rescattering. The goal is achieved first by utilizing general arguments provided by Landau equations, and second by applying a modern scattering formalism with explicit two- and three-body unitarity.
\end{abstract}
\maketitle%
\tableofcontents

\section{Introduction}
\label{sec:introduction}
The resonance spectrum of QCD is an emergent feature of the theory, manifesting its non-perturbative nature. Resonance parameters are encoded in the analytic structure of the transition amplitudes, i.e., in the poles and residues on the unphysical Riemann sheets in the complex energy plane. For a recent review on the subject, see e.g., Ref.~\cite{Mai:2022eur} (resonances in the presence of anomalous thresholds are considered, e.g., in Refs.~\cite{Lutz:2015lca,Lutz:2018kaz,Korpa:2022voo}). Resonances are usually classified as standard QCD resonances and exotic states. The former refers to the states that can be understood within the traditional quark model~\cite{Richard:2012xw} or self-consistent Dyson-Schwinger approaches~\cite{Eichmann:2016yit, Eichmann:2016hgl, Yin:2019bxe}, both based on the quark-antiquark and three-quark picture for the mesons and baryons, respectively. Exotic states are the ones that do not fit this picture naturally. These are, for example, tetraquarks, pentaquarks, hadronic molecules, etc.,~\cite{Maiani:2004vq, Heupel:2012ua,LHCb:2015yax} that emerge in the non-perturbative interactions of mesons and baryons~\cite{Bruns:2010sv, Lutz:2001yb, Oset:2016lyh, Mai:2012dt, Mai:2014xna}. The distinction between these categories is not always unambiguous for given quantum numbers, see e.g., Ref.~\cite{Guo:2015umn}. Furthermore, there are kinematic effects in scattering that can lead to structures in the amplitudes, cross sections and line-shapes. One example is the threshold cusp, which is directly visible in the data (see e.g., Ref.~\cite{Colangelo:2006va,Gasser:2011ju}) or in the partial-wave analysis~\cite{Briscoe:2023gmb}. A similar effect is given by the complex threshold openings, in which one or both of the scattered ``particles'' is a resonant hadronic subsystem itself, also referred to as an isobar~\cite{Aitchison:1964rwb,Ceci:2011ae}. One may observe this phenomenon at the energy at which a resonance and a spectator particle can go on-shell. The latter statement has to be treated with caution, and is applicable in the case of a narrow resonance only, because resonances, strictly speaking, are not asymptotic states. Closely related to this situation is the kinematic configuration when the isobar decays, and one of its decay products is absorbed by the spectator. If the resonant isobar, the spectator and the exchanged decay product of the isobar can simultaneously go on-shell, the so-called triangle singularity emerges~\cite{Bayar:2016ftu, Guo:2019twa, Isken:2023xfo, Debastiani:2018xoi}. This singularity has been conjectured for a plethora of amplitude structures~\cite{Dai:2018hqb, Dai:2018rra, Liang:2019jtr, Jing:2019cbw, Du:2021zdg, Duan:2023dky, Wang:2016dtb, Nakamura:2023obk, Zhang:2024dth, Achasov:2022onn, Nakamura:2023hbt}.

While the resonance spectrum provides direct access to QCD dynamics, the understanding of the kinematic structures is equally important to prevent misidentifications of bumps in experimental data as resonances, and to get insight into hadron dynamics. A prominent example concerns the structure in the P-wave $f_0 \pi^-$ line shape with the $a_1(1260)$ quantum numbers that is observed at energies well above the $a_1(1260)$ resonance~\cite{COMPASS:2015kdx, Rabusov:2023tna, Mikhasenko:2015oxp} in the $\pi^- \pi^+ \pi^-$ final state. Two scenarios exist to explain this excess: an excited state of the $a_1(1260)$-resonance, the so-called $a_1(1420)$, or a kinematic singularity arising in the $K^* K \rightarrow f_0 \pi$ rescattering through a $K$ exchange, see e.g., Refs.~\cite{Mikhasenko:2015oxp, Aceti:2016yeb, Guo:2019twa, COMPASS:2020yhb}. In the following, we put the latter scenario under a renewed scrutiny by first using Landau equations, and later the unitary three-body framework~\cite{Mai:2017vot, Sadasivan:2021emk} referred to as IVU (Infinite Volume Unitarity). Formally, this approach represents a re-formulation of the well-known Faddeev three-body equations~\cite{Faddeev:1960su} with a separable two-body input and a three-body force in the field-theoretical language with a special emphasis on two- and three-body unitarity. This framework was extended to finite volume (FVU)~\cite{Mai:2017bge, Mai:2021lwb}, and generalized later to coupled $\pi\rho$ channels in S- and D-partial waves~\cite{Sadasivan:2021emk,Mai:2021nul}, which allowed, for the first time, an extraction of a resonance pole in the three-particle sector from the experimental/lattice input. This method was adapted recently to extract the resonance parameters of the $\omega$-meson from lattice QCD, extrapolating it also to the physical point, see Ref.~\cite{Yan:2024gwp}. For the discussion of the methods of solution of the integral equations and the extraction of resonance poles, see e.g., Refs.~\cite{Doring:2009yv, Suzuki:2009nj, Lutz:2015lca, Jackura:2020bsk, Sadasivan:2021emk, Dawid:2023jrj, Dawid:2023kxu, Nakamura:2023obk, Garofalo:2022pux}. Three-body rescattering effects for the $\pi\rho$ system can also be studied using Khuri-Treiman equations with the two-body input in the form of Omn\`es-functions, as demonstrated recently~\cite{Stamen:2022eda}. For a discussion of the dispersion representations for the triangle diagram, see Ref.~\cite{Lucha:2006vc}.

In this work, we carry out a semi-quantitative study of the problem in question by making use of the multichannel IVU framework that includes the $K^*K$ and the $f_0\pi$ channels explicitly. On the one hand, the present study provides a test of the relevance of rescattering effects. On the other hand, it sets a stage for future analysis of experimental data that aims to quantify both (kinematic vs. resonance) effects reliably. This is carried out through the means of a framework which allows describing the decay process quantitatively~\cite{Fengprep}. However, in this paper, in order to make the discussion more transparent, a toy model is utilized which assumes the $K^*$ to be spinless, whereas other particles involved in the rescattering process, i.e., $K$, $f_0$, $\pi$ are innately spinless. Masses of all involved particles are taken equal to their physical values~\cite{ParticleDataGroup:2022pth}. Furthermore, we only consider the isobars and spectators in relative S-wave, and consequently the hypothetical $a_1(1420)$ meson or, better, the source with the quantum numbers of the $a_1(1260)$, is also spinless. In the following, we refer to this simply as the $a_1$, and no confusion should arise in this regard. The main motivation of this mock-up system roots in the fact that the triangle singularity arises entirely from the denominators of the relevant internal propagators, and is unaffected by the spins of the particles involved. It should also be pointed out that the study of the effect of final-state interactions for different physical systems has been carried out recently, see e.g., Refs.~\cite{Nakamura:2023hbt, Nakamura:2023obk, Zhang:2024dth, Liu:2024uxn,JPAC:2021rxu}. This paper, thus, represents a complementary piece of work carried out in a different setting, with a slightly more emphasis on the purely theoretical aspects of the problem. Hence, in the long run, the conceptual progress discussed here can contribute to reliable identification of structures as resonances or kinematic effects. This will become relevant in the future extensions of isobar analyses of three-body final states in the experiments for which, usually, no explicit exchange terms or coupled channels are included. See, for example, analyses at COMPASS~\cite{Ketzer:2019wmd, COMPASS:2018uzl}, Crystal Barrel~\cite{CrystalBarrel:2019zqh} (does include coupled channels), BESIII~\cite{BESIII:2023qgj}, Belle~\cite{Rabusov:2022woa}, or LHCb~\cite{LHCb:2022lja}.

This article is structured as follows. First, in \cref{sec:analytic-part}, we discuss the generalization of the Landau equations to the case when an infinite set of rescattering terms is included in the final state. In particular, we want to ensure that the inclusion of ladder diagrams neither changes the position of the existing triangle singularity, nor new ones appear. Then, in \cref{sec:IVU} we explicitly calculate the line shapes in a mock-up model resembling the main analytical features the $a_1$ system. This is based on the IVU approach and, thus, includes all the two- and three-body final-state rescattering terms and respects unitarity exactly. There, we also discuss the implementation of the formalism and compare explicitly various solution techniques of the three-body integral equations, see \cref{subsec:implementation}. Quantitative comparison is carried out and discussed in \cref{subsec:results}. Conclusions  of the paper and outlook for future work are summarized in \cref{sec:conclusions}.

\section{Singularities of the ladder diagrams}
\label{sec:analytic-part}
When the masses of the decay products as well as the particles running in the loop obey certain conditions, the triangle diagram leads to a logarithmic singularity. The aim of this section is to find out, to which extent this statement remains true when this triangle diagram is dressed by an infinite number of diagrams corresponding to two and three-body final-state rescattering. In this section, we address this question by studying the singularities of the relevant $n$-loop Feynman diagrams, where $n$ is arbitrary, with the use of the Landau equations.

In the physical problem we consider here, one of the particles in the triangle diagram is unstable. The decay product recombines with a spectator forming a second isobar, which can decay and recombine again, and so on. This demonstrates that the triangle diagram is only a substructure in an infinite rescattering series. This complicates the analysis considerably. Here, we first consider the case when all the internal particles are treated as stable particles, and will briefly comment on the case of unstable particles at the end of this section. For the quantitative assessment of these effects through the 3-body unitary IVU formalism~\cite{Mai:2017vot}, see \cref{sec:IVU}.

In the stable particles case, one has to consider only the ladder diagrams (see below), where the analysis on the basis of Landau equations can be carried out for any number of loops. The Landau equations, which determine the singularities of an arbitrary Feynman integral in the external kinematic variables, are based on the observation that non-analyticities may arise in the integrals, when the singularities of the integrand cannot be circumvented by deforming the contour of integration. This typically happens, when one evaluates the integral at one of its endpoints, or when the structure of the singularities of the integrand is such that two singularities come from the opposite sides and pinch the integration contour, making deformation impossible~\cite{Bayar:2016ftu}. This may happen only for certain configurations of external momenta, which are found by solving the Landau equations. For a recent review, see Ref.~\cite{Guo:2019twa}.

Now, consider an arbitrary Feynman integral with $N$ internal propagators with momenta $q_i$ and $L$ loop momenta labeled by $k_j$. The former momenta are linear combinations of the latter and of the external momenta, with the coefficients $\pm 1$ (in addition, one can choose the momenta so that each loop momentum $k_j$ enters in all $q_i$ with the coefficient $+1$). The Landau equations~\cite{Landau:1959fi,Polkinghorne:1960cjb,Polkinghorne:1960udx} then read:
\begin{align}
    \label{eqn:landau1}
    \alpha_i(q_i^2-m_i^2) = 0\,,&\quad\quad i=1,\ldots,N,\\
    \label{eqn:landau2}
        \sum_{i\, \in\, \textrm{loop}\,j} \alpha_i q_i^{\mu}(k_j) = 0\,,&\quad\quad j=1,\ldots, L,
  \end{align}
where $m_i$ is the mass in the corresponding propagator and $\alpha_i$ denotes the Feynman parameter for the internal line $i$. Note that the second sum runs over all internal lines that contain a given loop momentum (that is, ${dq_i^\mu}/{dk_j^\nu}=\delta^\mu_\nu\neq 0$). Hence, there are $N$ equations corresponding to \cref{eqn:landau1} and $L$ equations corresponding to \cref{eqn:landau2}. The first condition translates to either the internal propagators going on mass shell or the corresponding Feynman parameter vanishing. The latter case occurs when the internal propagator does not contribute to the integral. One can identify the resulting Feynman diagram as a graph with the corresponding propagator \emph{contracted}. When all the internal propagators go on mass shell, the corresponding singularity is called the leading Landau singularity. When one or more of the Feynman parameters vanish, the corresponding singularity is called the subleading Landau singularity. The second condition translates to the $4$-momenta of the propagators being co-planar. Furthermore, it can be shown that all the Feynman parameters must obey the condition $\alpha_i \geq 0$ for the singularity to be present on the physical sheet~\cite{Coleman:1965xm}.

For the leading Landau singularity of the triangle diagram, the Landau equations can be written as a matrix equation by taking the dot product of \cref{eqn:landau2} with every $q_i$. The resulting equation reads:
\begin{equation}
    \label{eqn:trianglematrix}
    Q \bm{\alpha} = \bm{0},
\end{equation}
where the elements of the matrix $Q$ are given by $Q_{ij} = q_i \cdot q_j$ and the vector $\bm{\alpha}$ denotes $(\alpha_1,\alpha_2,\ldots)^T$. The non-trivial solutions of \cref{eqn:trianglematrix} can be obtained by requiring the determinant to be zero, i.e., $\det Q = 0$. Noting again that, for the leading Landau singularity, \cref{eqn:landau1} implies that the internal propagators are on-shell, i.e., $q_i^2 = m_i^2$, the solutions to the determinant equation can be written entirely in terms of the invariant masses of the external particles.
\begin{figure}[t]
    \begin{center}
        \includegraphics[height=2.5cm]{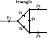}
        \hspace{0.5cm}
        \includegraphics[height=2.5cm]{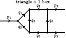}
        \hspace{0.5cm}
        \includegraphics[height=2.5cm]{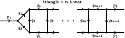}
        \caption{The first diagram shows the decay of a particle (with momentum $p_1$) into a pair with momenta $p_2,p_3$, which will be identified as the $f_0$ and $\pi$ mesons, respectively. Here, $q_{1,2,3}$ denote the internal momenta for the particles with the masses $m_{1,2,3}$ running in the loop. Adding one loop to the triangle diagram corresponds to a diagram with final-state rescattering, as in the second diagram. Similarly, an arbitrary number of interactions in the final states can be considered, as shown in the last diagram with $n$ boxes attached to the original triangle diagram. This naming convention of the diagrams is followed throughout.}
        \label{fig:landau-triangle}
    \end{center}
\end{figure}

We are interested in a two-particle decay process where a (hypothetical) meson with the momentum $p_1$ decays into the $f_0$ and $\pi$ mesons with momenta $p_2$ and $p_3$, respectively. The external momenta obey the conditions given by $p_2^2=\sigma$ and $p_3^2=m_\pi^2$, respectively, where $\sigma$ is the square of the invariant mass of the $f_0$ isobar. Bearing in mind that $f_0$ is unstable, meaning that its mass is not exactly fixed, we wish to use this freedom to scan a certain interval in $\sigma$, looking for singularities. Similarly, the momentum of the decaying particle obeys the relation $p_1^2=s$, and we scan again an interval in this variable around the singular point.

Consider the simple triangle diagram first, depicted in~\cref{fig:landau-triangle}. In this case, the vanishing determinant associated with the Landau equations can be written down in terms of $s$ and $\sigma$: 
\begin{equation}
    \label{eqn:landaudeterminantsimple}
    y_{12}^2 + y_{23}^2 + r^2 - 2r\, y_{12} y_{23} - 1 = 0,
\end{equation}
where $y_{12} = (m_1^2 + m_2^2 - s)/(2m_1m_2)$, $y_{23} = (m_2^2 + m_3^2 - \sigma)/(2m_2 m_3)$ and $r=(m_1^2 + m_3^2 - m_\pi^2)/(2m_1 m_3)$. The triangle singularity can occur on the physical boundary if and only if the invariant masses are greater than or equal to the two-body thresholds of the particles attached to them, i.e., $s\geq(m_1+m_2)^2$ and $\sigma\geq(m_2 + m_3)^2$. This lower bound on one of the invariant masses imposes an upper bound on the other invariant mass through \cref{eqn:landaudeterminantsimple}. Equivalently, one can also derive these conditions in terms of bounds on $m_1$:
\begin{align}
    m_{1,{\sf low}}<m_1<m_{1,{\sf up}}\, ,\quad\quad
    m_{1,{\sf low}}^2=\frac{sm_3+m_\pi^2m_2}{m_2+m_3}-m_2m_3\, ,\quad\quad
    m_{1,{\sf up}}^2=(\sqrt{s}-m_2)^2\, .
\end{align}
Again, by making use of \cref{eqn:landaudeterminantsimple}, in terms of $\sigma$ and $s$ this translates to
\begin{align}
    \label{eqn:landau-triangle1}
    (m_2+m_3)^2&<\sigma<m_2^2+m_3^2+\frac{m_2}{m_1}(m_1^2+m_3^2-m_\pi^2)\, ,
    \\[2mm] \label{eqn:landau-triangle2}
    (m_1+m_2)^2&<s<m_1^2+m_2^2+\frac{m_2}{m_3}(m_1^2+m_3^2-m_\pi^2)\, .
\end{align}
For the details of the derivation, we refer the reader to Ref.~\cite{Guo:2019twa}.

In addition, the triangle graph has subleading singularities in the incoming and outgoing invariant masses, which are exactly the corresponding two-body thresholds $s = (m_1 + m_2)^2$ and $\sigma = (m_2 + m_3)^2$. The two-body threshold in $s$ corresponds to the case when the Feynman parameter $\alpha_3$ vanishes, leading to a graph with a contracted $m_3$ line. The two-body threshold in $\sigma$ corresponds to the case when $\alpha_1$ vanishes, leading to a graph with contracted $m_1$ line. The latter point means that particle 3 and particle 1 do not contribute to the Feynman integrals in the respective cases. When the relevant amplitudes are evaluated for the incoming and outgoing invariant masses outside the interval given in \cref{eqn:landau-triangle1} and \cref{eqn:landau-triangle2}, the subleading singularities are still present and manifest themselves as cusps.

Scrutinizing the effect of final-state interactions, we add ladder diagrams to the triangle diagram. The diagram with a single loop added is shown in the second subfigure of~\cref{fig:landau-triangle}. In this case, the Landau equations that determine the leading singularity are given by
\begin{equation}
    \label{eqn:house-landau1}
    q_i^2 = m_i^2,
\end{equation}
which is obtained from \cref{eqn:landau-triangle1}, and ensures that all the internal propagators are on-shell. Furthermore,
\begin{align}
    \label{eqn:house-landau2}
    Q_1 \boldsymbol{\alpha}&= \bm{0} \\
    \label{eqn:house-landau3}
    Q_2 \boldsymbol{\alpha} &= \bm{0},
\end{align}
where the matrix elements of the block-diagonal matrix $Q_1$ are given by $(Q_1)_{ij} = q_i \cdot q_j$ for $i,j =1,2,3$, resulting from applying \cref{eqn:landau2} to the first loop momentum. Analogously, the matrix elements of $Q_2$ are given by $(Q_2)_{ij} = q_i \cdot q_j$ for $i,j=3,4,5,6$. Now, one needs to find a simultaneous solution to \cref{eqn:house-landau2} and \cref{eqn:house-landau3}, and subsequently verify that all corresponding Feynman parameters are positive.

First, we consider finding a simultaneous solution to \cref{eqn:house-landau2} and \cref{eqn:house-landau3}, which reduces to finding the simultaneous solution to ${\rm det}\,Q_1={\rm det}\,Q_2=0$. The first determinant can be written down entirely in terms of $s$, owing to the fact that every other particle involved here is on-shell by virtue of \cref{eqn:house-landau1}. The second determinant depends on $s$ and, additionally on the Mandelstam variable $t = (p_2 - q_2)^2$ with $p_2^2 = \sigma$. One can express $t$ in terms of $s$ and $\sigma$. Solving the first equation with respect to $s$, drawing then all possible diagrams with $K^*, K^\pm, f_0, \pi^\pm$, running in the loops, and solving the second equation with respect to $\sigma$ (for a given $s$) in order to find the leading Landau singularity, we arrive at complex-valued solutions for $\sigma$ for any above choice of the particle masses in the loops. Thus, the two-loop ($\mbox{triangle}+1\,\mbox{box}$) diagram, depicted in \cref{fig:landau-triangle}, does not possess the leading Landau singularity on the real axis. This also eliminates the possibility of the leading Landau singularity in two rescatterings and beyond, since any additional rescattering will add another matrix equation that requires a simultaneous solution with all preceding matrix equations. Hence, from now on, we can concentrate on analyzing the subleading singularities.

\begin{figure}[t]
    \centering
    \includegraphics[width=0.7\linewidth]{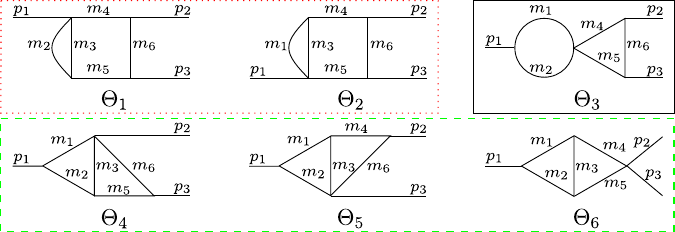}
    \caption{All possible single contractions of the $\mbox{triangle}+1\,\mbox{box}$ graph. The singular triangle graph is in the black box. The notation $\Theta_i$ stands for a graph in which the $i$-th internal line is contracted (we use this notation consistently throughout this paper.)}
    \label{fig:1contractions}
\end{figure}

For the subleading singularities, we need to consider all possible contractions. There are $\binom{6}{1} = 6$ single contractions, see \cref{fig:1contractions}. Note that, following \cref{eqn:landau1}, we have used the on-shell masses to label the internal propagators instead of their momenta. The graphs inside the dotted red box require a simultaneous solution to a bubble and a square graph. The graphs inside the dashed green box require a simultaneous solution to the two triangle graphs. For the scenario considered in this work, no combinations of masses exists such that the above equations are fulfilled simultaneously. 
The graph inside the black box ($\Theta_3$) factorizes into a bubble and a triangle graph, i.e., there is no propagator that depends on more than one loop momentum. This case requires independent solutions to the two subgraphs. The latter of which results in a triangle singularity for the relevant propagators.

\begin{figure}[t]
    \centering
    \includegraphics[width=\linewidth]{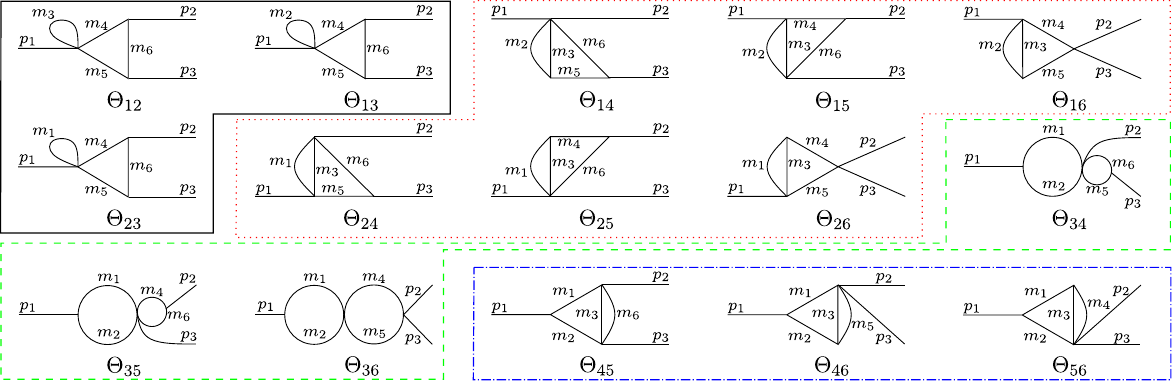}
    \caption{All possible double contractions of the $\mbox{triangle}+1\,\mbox{box}$ graph. The singular triangle graphs are in the black box.}
    \label{fig:2contractions}
\end{figure}

Next, there are $\binom{6}{2} = 15$ double contractions, see \cref{fig:2contractions}. The graphs inside the dotted red box and the dot-dashed blue box require a simultaneous solution to a bubble and a triangle graph. In this case, for the various combinations of the propagators in this analysis, a solution does not exist. The graphs inside the dashed green box factorize into two bubble graphs. This case results in the relevant two-body thresholds. The graphs inside the black dashed box factorize into a tadpole graph and a triangle graph. Tadpole graphs are trivial and can be ignored when analyzing the Landau singularities. We are then left with the triangle graph that reproduces the original triangle singularity.

\begin{figure}[t]
    \centering
    \includegraphics[width=\linewidth]{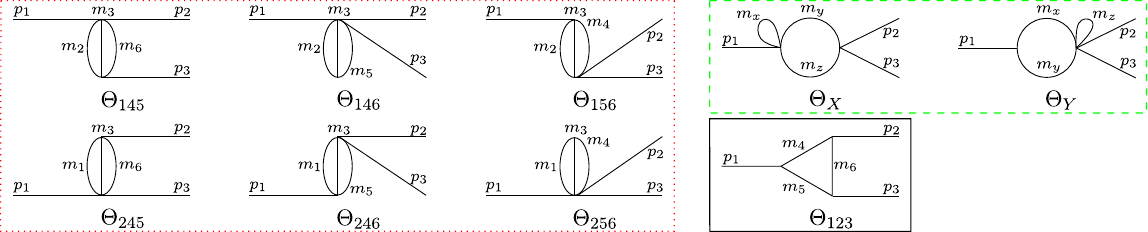}
    \caption{Triple contractions of the $\mbox{triangle}+1\,\mbox{box}$ graph. The two graphs in the dashed green box correspond to the different combinations of the contracted propagators, see the text for details. The singular triangle graph is in the black box.}
    \label{fig:3contractions}
\end{figure}

Next, there are $\binom{6}{3} = 20$ triple contractions, see \cref{fig:3contractions}, where we have omitted showing all possible permutations of different masses for the graphs in the dashed green box explicitly to be concise. The ``coffee bean'' graphs inside the dotted red box require a simultaneous solution to two bubble graphs. In this case, for the various combinations of the propagators considered in this analysis, a solution does not exist. The graphs inside the dashed green box factorize into a tadpole graph and a bubble graph, which results in the relevant two-body thresholds. Additionally, $9$ graphs --- $\{\Theta_{124}, \Theta_{125}, \Theta_{126}, \Theta_{134}, \Theta_{135}, \Theta_{136}, \Theta_{234}, \Theta_{235}, \Theta_{236}\}$ --- are denoted as $\Theta_X$, and $4$ graphs --- $\{\Theta_{345}, \Theta_{346}, \Theta_{356}, \Theta_{456}\}$ --- are denoted as $\Theta_Y$. The graph inside the black box is the triangle graph, which again results in a triangle singularity for the relevant propagators.
Finally, there are $\binom{6}{4} = 15$ quadruple contractions, all of which lead to bubble graphs. And more contractions lead to trivial graphs. These observations are essential, when we consider an arbitrary number of final-state interactions.

\begin{figure}[t]
    \begin{center}
        \includegraphics[width=10cm]{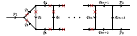}
        \caption{In the $\mbox{triangle}+n\,\mbox{boxes}$ diagram, the pertinent subleading singularity arises when every propagator except the final three in the ladder is contracted, for example.}
        \label{fig:nbox}
    \end{center}
\end{figure}

For a graph with $n+1$ rescatterings (\cref{fig:nbox}), first note that there are contractions that lead to the factorized graphs with non-factorizable sub-blocks. If we show that there are no solutions to an arbitrary non-factorizable graph, then obviously there are also no solutions to graphs made up of non-factorizable subgraphs. Therefore, we concentrate on the non-factorizable subgraphs of the $\mbox{triangle}+n\,\mbox{boxes}$ diagram. Now, for any non-factorizable subgraph, one needs to find a simultaneous solution to $n+1$ determinant equations. For the first two determinant equations that correspond to the triangle and the utmost left box, we already know that there are no singular subgraphs other than the graphs that lead to two-body thresholds or the original triangle singularity. However, now the singular diagrams $\Theta_3,\Theta_{12},\Theta_{13},\Theta_{23},\Theta_{123}$, shown in Figs.~(\ref{fig:1contractions}, \ref{fig:2contractions} and \ref{fig:3contractions}), are embedded in the diagrams with more boxes. Consequently, their external legs corresponding to the outgoing momenta $p_2,p_3$ turn into the integration momenta and the singularity is washed out. Hence, one has to look for the subleading singularities again, considering all possible contractions that involve the second box to the left. Proceeding this way and eliminating all the boxes one after another, we come to the conclusion that a generic $n$-box diagram reproduces original triangle singularity as a subleading singularity --- the pertinent contracted diagram is factorized into the original triangle loop and any number of bubble and tadpole diagrams.

Before moving ahead with the investigation of the triangle singularities through the three-body unitary formalism, we wish to remark on three issues. Firstly, final-state interactions, in general, are not described solely in terms of the ladder diagrams. In addition, one has four-point contact interactions, e.g.,  $f_0\pi\to f_0\pi$ scattering through a local four-point vertex. In the three-particle equations in \cref{sec:IVU}, such an interaction is needed to ensure the cutoff-independence of the physical amplitude. We, however, note that, inserting these four-point vertices\footnote{In case there is a nearby three-particle resonance in the $s$-channel, the derivative expansion of the local isobar-spectator interaction will possess a very small radius of convergence. An obvious cure to this problem is to introduce an $s$-dependent local interaction, and assume that these couplings contain a simple pole in $s$. Our argument remains in place in this case as well.} into the ladder diagrams, one arrives at diagrams whose topology is similar to the ladder diagrams with the contracted exchange particle propagator. Thus, no change of the singularity structure is expected from this.

Secondly, it should be noted that all considered diagrams have exactly the same structure (strength) of singularity, since they contain the same singular triangle diagram. This means that the singularity stays exactly at the same place after resumming all the boxes from final-state interactions. Note that this is not the case, for example, when one considers dressing the resonance or a bound-state pole with final-state interactions. In the perturbation theory for the propagator of a bound state (bubble sum), one has diagrams containing $n$ unperturbed propagators and $n-1$ bubbles (self-energy diagrams). A generic contribution in perturbation series is written as $(s-s_0)^{-n}B(s)^{n-1}$, where $s_0$ denotes the unperturbed value of the pole position and $B(s)$ stands for a single bubble. In other words, we have poles of different order in different terms of the perturbative series. Furthermore, summing up the infinite series, it can be seen that the pole position is shifted due to final-state interactions, $s_0\to s_0+\delta s$, where the shift is determined from the equation $\delta s=B(s_0+\delta s)$. This phenomenon does not occur in the case of triangle singularity, because all the terms in the perturbative series have exactly the same type of singularity.

Finally, another challenging problem is related to the fact that at least one of the particles in the triangle is unstable (see \cref{sec:IVU}). For the one-loop triangle diagram, this problem has been investigated in detail, e.g., in Ref.~\cite{Aitchison:1964rwb}. To briefly summarize, in this paper the K\"allen-Lehmann representation for the propagator of the unstable particle was used. It was assumed that, instead of the real axis, the propagator has a pole on the second Riemann sheet close to the cut, which is equivalent to assuming the spectral function having a pole exactly at the same position (on the first sheet). The full result for the triangle diagram corresponding to summing all the bubbles in the internal line is given by a convolution of this spectral function with the triangle diagram, in which one integrates over the mass assigned to this internal line. It was argued then that, if the singularity of the spectral function lies very close to the real axis, the position of the singularity in the final expression is very close to the one obtained from a triangle diagram with the resonance mass parameter replaced by the complex resonance pole position. Note also that, in the context of the two-body scattering, the described behavior is known under the name of ``woolly cusps''~\cite{PhysRev.126.360} or complex threshold openings, since the resonance mass is not a sharp value but has a finite width moving thresholds into the complex plane~\cite{Doring:2009yv, Ceci:2011ae}.

Extending this argument to the series of ladder diagrams considered in the present section, one needs to replace each mass of the unstable particles by their complex resonance pole positions. This will move the triangle singularity away from the real axis. By using numerical integration, we have explicitly checked that, in case of the two-loop diagram shown in \cref{fig:landau-triangle}, the final-state interaction has small effect on the position and shape of the leading-order triangle singularity. In the next section, we provide a more rigorous check of this statement, using the IVU framework.

\section{Singularities in the IVU framework}
\label{sec:IVU}
\subsection{Formalism}
\label{sec:IVU-Formalism}

In order to avoid the clutter of indices, we first briefly discuss the basis notions of the IVU-approach~\cite{Mai:2017vot} in the one-channel case. In this approach, one rewrites the three-body amplitude as an infinite set of diagrams describing interactions between a two-body cluster (sometimes referred to as the isobar or the dimer) and a third particle (sometimes referred to as the spectator or the bachelor) through particle exchange, as well as through short-range interactions. The isobar is represented through a bubble sum, while the spectator is on mass shell. The amplitude is decomposed into a fully connected and disconnected parts. Since the disconnected piece does not lead to the relevant triangle singularity, it is sufficient to discuss only the connected piece in the following. This piece can be expressed via the isobar-spectator scattering amplitude which obeys the following three-dimensional integral equation:
\begin{equation}
    \label{eqn:bethesalpeter-T}
    T(\sqrt{s},\bm{q},\bm{p}) = \left( B(\sqrt{s},\bm{q},\bm{p})+C(\sqrt{s},\bm{q},\bm{p}) \right) + \int \frac{d^3l}{(2\pi)^3} \frac{1}{2 E_{\bm{l}}} T(\sqrt{s},\bm{q},\bm{l}) \tau(\sigma(\bm{l})) \left(B(\sqrt{s},\bm{l},\bm{p})+C(\sqrt{s},\bm{l},\bm{p})\right).
\end{equation}
Here, $\bm{q}$ and $\bm{p}$ are the outgoing and incoming on-shell spectator $3$-momenta, respectively, and $\bm{l}$ is the spectator 3-momentum in the intermediate state. Furthermore, $\sqrt{s}$ is the $3$-body invariant mass, which, in the center-of-mass (CM) frame that we work in, is just the total energy. Finally, $\sqrt{\sigma(\bm{l})}$ is the invariant mass of the $2$-body subsystem (the isobar), written in terms of the spectator momentum ($E_{\bm{l}}=\sqrt{\bm{l}^2+m_{\bm{l}}^2}$), that is, $\sigma(\bm{l}) = (\sqrt{s}-E_{\bm{l}},-\bm{l})^2 = s + m_l^2 - 2\sqrt{s}E_{\bm{l}}$. Here and in the following we denote the mass of the particle carrying momentum $x$ as $m_x$ as well as $x=|\bm{x}|$ for any considered three-momentum $\bm{x}$. The integral equation contains the isobar propagator $\tau$. The kernel of the equation consists of the one-particle exchange term $B$ and the isobar-spectator contact term $C$, which is discussed below. To access the line-shapes, we integrate the isobar-spectator amplitude~\cref{eqn:bethesalpeter-T} with respect to the 3-momentum of the incoming state as follows:
\begin{equation}
    \label{eqn:connected-T}
    \Gamma(\sqrt{s},\bm{q}) = \int \frac{d^3 \bm{p}}{(2\pi)^3} \frac{1}{2 E_{\bm{p}}}\, 
T(\sqrt{s},\bm{q},\bm{p})\,
 \tau(\sigma(\bm{p}))\,
 D(\sqrt{s},\bm{p}).
\end{equation}
Here, $D$ is the scalar vertex associated with the dissociation of the $a_1$ into an isobar and a spectator. This vertex, along with the contact term $C$, contains all free parameters at our disposal which can be used to describe the three-body scattering in the channel with given quantum numbers. In principle, these parameters should be fit to experimental data. This is set aside for a future study, since the dynamical effect of triangle singularity shows little sensitivity to the details of the short-range interactions. In particular, in the following, we neglect all momentum dependence in $D$, taking it to be constant. With this, $\Gamma$ gives the amplitude, associated with $a_1$ decaying into an isobar and a spectator that also includes all final-state interactions. We note that the physical line shape can be obtained from this quantity by adding the disconnected contribution, and then multiplying everything with the final isobar propagator and the decay vertex into asymptotically stable states (e.g. $3\pi$). The details of this procedure are given in Ref.~\cite{Sadasivan:2020syi} but are of no relevance here since $\Gamma$ contains all the relevant singularities and rescattering diagrams.

Coming back to the ingredients of \cref{eqn:bethesalpeter-T}, we first note that the isobar-spectator contact term $C$ (three-body force), which ensures cutoff-independence of the physical observables obtained from the solution of this integral equation~\cite{Brett:2021wyd}, cannot produce the triangle singularity. For this reason, we first completely neglect it, and enforce a sharp cutoff (which is equivalent to the choice of the renormalization prescription). Next, however, we will relax this assumption allowing it to have a pole in Mandelstam $s$ if demanded by the data, see the discussion in Ref.~\cite{Sadasivan:2021emk}. This pole, however, will be irrelevant since it lies far from the energy region that we are interested in.

Next, we consider the isobar propagator $\tau$ in \cref{eqn:bethesalpeter-T}, which encodes the two-body interaction. Explicitly, for distinguishable particles in the self-energy equation, it is given by
\begin{align}
    \label{eqn:tau-term}
    \tau(\sigma(\bm{p})) 
    &=
    \frac{1}{\sigma(\bm{p}) - m_{\text{bare}}^2 - g^2\Sigma(\sigma(\bm{p}))+i\epsilon} 
    \,,\\
    \label{eqn:tau-se}
    \Sigma(\sigma(\bm{p})) 
    &= 
    \int \frac{d^3 \bm{k}}{(2\pi)^3} \frac{1}{2E_1(\bm{k})E_2(\bm{k})} \frac{\sigma(\bm{p})}{(E_1(\bm{k})+E_2(\bm{k}))^2} \frac{(E_1(\bm{k})+E_2(\bm{k}))}{\sigma(\bm{p}) - (E_1(\bm{k})+E_2(\bm{k}))^2+i\epsilon}\,.
\end{align}
Here, the two-body self-energy $\Sigma$ is written down in a once-subtracted form to ensure convergence of the integral (note also that the self-energy for indistinguishable particles like two pions carries another factor of $1/2$). This expression is then evaluated in the isobar CM frame. 
The unknown constants $(g,m_{\text{bare}})$ are determined through the physical values (the mass $m_{\text{phys}}$ and the width $\Gamma$ of a resonance), which for narrow resonances simply yields
\begin{subequations}
\begin{align}
  m_{\text{phys}}^2 &=m_{\text{bare}}^2+g^2\Re{(\Sigma(m_{\text{phys}}^2))_{\sf II}}\,, \\
m_{\text{phys}} \Gamma&=-g^2\Im{(\Sigma(m_{\text{phys}}^2))_{\sf II}}\,,
\end{align}
\end{subequations}
where the subscript ${\sf II}$ denotes that the value of the self-energy is taken on the second sheet. It is also possible to match the isobar propagator to a more realistic two-body scattering amplitude, including, for example, constraints from Chiral Perturbation Theory, for details see Refs.~\cite{Culver:2019vvu,Alexandru:2020xqf,Mai:2021nul,Mai:2018djl,Feng:2024wyg}.

Finally, the one-particle exchange diagram $B$ is given by
\begin{equation}
    \label{eqn:b-term}
    B(\sqrt{s},\bm{q},\bm{p}) =\frac{g^2}{2E_{\bm{q}+\bm{p}}(\sqrt{s}-E_{\bm{q}}-E_{\bm{p}}-E_{\bm{q+p}}+i\epsilon)}\,,
\end{equation}
where the $g$'s denote the scalar-isobar dissociation vertex, and the angle dependence enters through the energy of the exchanged particle $E_{\bm{q+p}}=\sqrt{(\bm{q}+\bm{p})^2+m_{pq}^2} =\sqrt{\bm{q}^2 + \bm{p}^2 + 2 \bm{q} \cdot \bm{p}+m_{pq}^2}$. Note that $m_{pq}$ denotes the mass of the exchanged particle. The analytic form of $B$ is fixed through three-body unitarity, as shown in Ref.~\cite{Mai:2017vot} (in particular, the quantity $g$ in \cref{eqn:tau-se}), and \cref{eqn:b-term} must be the same to ensure the two- and three-body unitarity of the solutions of the integral equation). This form leads to the non-trivial cut structures in the kernel of the integral equation that manifests itself after the partial-wave expansion. More technical details about the partial-wave expansion in three-body equations (general case) can be found in Refs.~\cite{Sadasivan:2020syi,Sadasivan:2021emk}.\footnote{It is also worth mentioning that, using partial-wave expansion in a finite-volume version of the formalism leads to numerical instabilities for selected values of the CM energy~\cite{Doring:2018xxx}. In this case, it is more convenient to work in the plane-wave basis and carry out the partial diagonalization of the 3-body quantization condition in various irreps of the octahedral group of the pertinent subgroups thereof~\cite{Doring:2018xxx,Mai:2017bge, Alexandru:2020xqf, Culver:2019vvu, Mai:2018djl}.} In our toy model where only S-wave short-range interactions are retained, the partial-wave expansion simplifies considerably and reduces to the expansion of the kernel $B$: 
\begin{equation}
    \label{eqn:b-integrated}
    B(\sqrt{s},q,p) \coloneqq \int d\Omega \,B(\sqrt{s},\bm{q},\bm{p}) = \frac{\pi g^2}{2pq} \log \left( \frac{\sqrt{s} - \sqrt{p^2 + m_p^2} - \sqrt{q^2 + m_q^2} - \sqrt{(p + q)^2 + m_{pq}^2}+i\epsilon}{\sqrt{s} - \sqrt{p^2 + m_p^2} - \sqrt{q^2 + m_q^2} - \sqrt{(p - q)^2 + m_{pq}^2}+i\epsilon} \right).
\end{equation}

\begin{figure}[!t]
    \includegraphics[height=7.5cm]{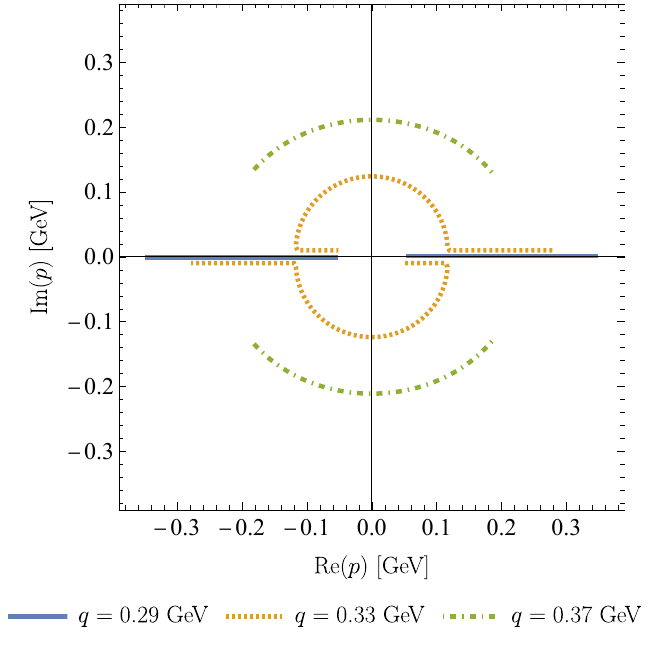}
    ~~~~
    \includegraphics[height=7.5cm]{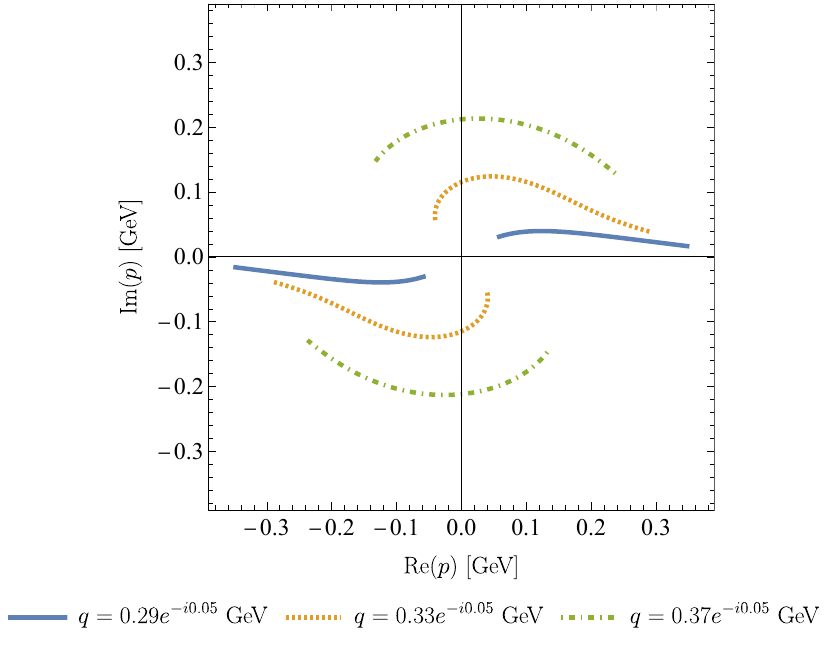}
    \caption{On the left, the singularities of the kernel for different {\em real} outgoing spectator $3$-momenta: three representative values are taken. On the right, the branch cuts for different {\em complex} outgoing spectator $3$-momenta. Both plots are shown for $\sqrt{s}=1.42~\textrm{GeV}$.}
    \label{fig:branch-cuts}
\end{figure}
The kernel given in \cref{eqn:b-integrated} contains logarithmic singularities. When evaluating the radial part of the integral given by \cref{eqn:connected-T}, one needs to account for the branch cuts associated with the logarithmic branch points. The integral can be numerically evaluated without any complications for invariant masses lying outside the Landau singularity regions given by \cref{eqn:landau-triangle1} and \cref{eqn:landau-triangle2} by simply numerically integrating along a deformed contour. However, this is no longer possible when the invariant mass lies within the Landau singularity region given in \cref{eqn:landau-triangle1} and \cref{eqn:landau-triangle2}. This complexity arises due to the circular branch cuts shown in \cref{fig:branch-cuts}. One way to circumvent this problem is to promote the in/outgoing spectator $3$-momentum to complex values, and then evaluate the integral numerically along the deformed contour, see again \cref{fig:branch-cuts}. After this, we still have to determine the amplitudes for real momenta, which is discussed in the next subsection.

We conclude this section by briefly considering the Born series of the decay amplitude $\Gamma$. This series can be schematically written as
\begin{equation}
    \Gamma = 
    \underbrace{~D \tau B~}_{\text{triangle}} +
    \underbrace{~D \tau B\tau B~}_{\text{triangle$+1$box}} +  
    \underbrace{~D \tau B\tau B\tau B~}_{\text{triangle$+2$boxes}}+\dots\,.
    \label{eq:series}
\end{equation}
Note also that the isobar propagator $\tau$ includes two-body rescattering effects to all orders. Below, we shall numerically evaluate the subleading term in this expansion and compare it with the full solution, demonstrating the fast convergence of the Born series.

\subsection{Implementation}
\label{subsec:implementation}
Moving forward with the Landau singularities in the case of the $a_1$, we discuss below the implementation of the IVU formalism, and also the numerical values of the relevant parameters in the formalism.
The problem at hand naturally requires at least two channels, without which there is no possibility to generate triangle singularity. These consist of the isobar-spectator pairs $K^*K$ and $f_0\pi$, which we refer to in the following as channel 1 and channel 2. In a physical system, $K^*K$ stands for isospin- and $G$-parity projected combinations of kaons with zero overall strangeness and negative electric charge. A quantitative analysis including isospin factors, spin structure and comparison to the experimental data is relegated to a future work as discussed before. Here, within the framework of our toy model, we include the channel with neutral $K^*$ only, which considerably simplifies the bookkeeping of diagrams but does not affect the singularity structure of the amplitude in the energy region we are interested. 
\begin{figure}[!t]
    \centering
    \includegraphics[width=0.7\linewidth]{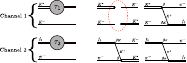}
    \caption{On the left, the two channels considered in this work, along with the relevant couplings. Only neutral $K^*$ mesons are included in our toy model. Furthermore, $\tau$ denotes the isobar propagator --- see \cref{eqn:tau-term} for $\tau_1$ and \cref{eqn:tau-term-mod} for $\tau_2$. On the right, the allowed transitions between the channels. Conservation of strangeness prohibits a transition from channel 1 to itself. This is indicated by a red dashed oval. However, the remaining transitions are allowed.}
    \label{fig:coupledchannel-transitions}
\end{figure}

The quantity $B$ corresponds to one-particle exchange between the channels. In our toy model, strangeness conservation prohibits a transition from channel 1 to channel 1. However, all other possible transitions are allowed as depicted in \cref{fig:coupledchannel-transitions}. Three coupling constants $g$, $g_K$ and $g_\pi$, corresponding to the transitions $K^{*0}\to K^+ \pi^-$, $f_0\to K^+ K^-$ and $f_0\to\pi^+\pi^-$, respectively, are needed to construct the kernel of the equation. These coupling constants, along with the bare masses of respective isobars, are fixed through the physical masses and widths of the $K^*$ and $f_0$ mesons. In this matching, we use the experimental input~\cite{ParticleDataGroup:2022pth} $m_{f_0}=990\,\MeV$, $\Gamma_{f_0}=50\,\MeV$ and $m_{K^*}=892\,\MeV$, $\Gamma_{K^*}=50\,\MeV$. It should be noted that the width of $f_0$ has little influence on the triangle singularity, since it does not contribute to the singular triangle graph. However, larger values of the $K^*$ indeed smear out the triangle singularity, see also Ref.~\cite{Guo:2019twa}.

In the case of the $f_0$ isobar, there are two different couplings entering the pertinent propagator (cf. \cref{eqn:tau-term})
\begin{equation}
    \label{eqn:tau-term-mod}
    \tau(\sigma(p)) = \frac{1}{\sigma(p) - m_{\text{bare}}^2 - g_K^2\Sigma_K(\sigma(p)) - g_\pi^2\Sigma_\pi(\sigma(p))+i\epsilon}\,.
\end{equation}
Here, $\Sigma_K$ and $\Sigma_\pi$ denote charged kaon and pion loops, respectively. Furthermore, in order to fix three couplings $g$, $g_K$ and $g_\pi$ from two decay widths, we use the input $g_K/g_\pi = 4.21$ from $\textrm{BES}$~\cite{BES:2004twe} and $\textrm{BaBaR}$~\cite{BaBar:2006hyf} collaborations. Finally, the parameters of our toy model determined from the above input are
\begin{align}
    &m_{\text{bare}}(K^*)=0.902\,\GeV\,,
    \quad
    g=1.860\,\GeV,\\\nonumber
    &m_{\text{bare}}(f_0)=1.089\,\GeV\,,
    \quad
    g_\pi=1.397\,\GeV\,,
    \quad
    g_K=5.880\,\GeV\,.
\end{align}

\bigskip
What remains now is to solve the integrals given by \cref{eqn:bethesalpeter-T} and \cref{eqn:connected-T} to obtain the amplitudes we are interested in. \cref{eqn:bethesalpeter-T} is a Fredholm integral equation of the second kind, which can be solved through the resolvent formalism. Evaluating this equation for real values of spectator momenta leads to a singular resolvent, as discussed above. Hence, following seminal works~\cite{Hetherington:1965zza,Cahill:1971ddy}, we promote the outgoing spectator momenta to complex values. In more details, the solution strategy is as follows:
\begin{enumerate}
\item \emph{Find a contour on the complex momentum-plane for which the resolvent is non-singular.}
We use a hard cutoff of $\Lambda=1~\textrm{GeV}$ in the integrals over spectator momenta, whereas the self-energy integral, which enters the isobar propagator, does not have a cutoff.
Since we are considering two channels, all our amplitudes are $2\times 2$ matrices in the channel space, and we have to ensure that the deformed contour does not hit the singularities of the kernel in any channel. In the integral over spectator momentum, the choice of the contour, similar to Ref.~\cite{Sadasivan:2021emk}, is given by:
\begin{equation}\label{eq:path}
    f_{\text{SMC}}(t) = t + i a(1-e^{-t/b})(1-e^{(t-\Lambda)/b})\,,\quad\quad 0\leq t\leq\Lambda\,,
\end{equation}
where $a$ and $b$ are some real parameters and $\Lambda$ is the cutoff. We refer to this contour as the \textit{Spectator Momentum Contour (SMC)}. Crucially, this choice of the contour ensures that the integration path, indeed, approaches the real axis again at the point $l=\Lambda$. For the self-energy integral, the choice, again similar to Ref.~\cite{Sadasivan:2021emk}, is given by
\begin{equation}
    f_{\text{SEC}}(t) = t + \frac{i}{2} c \tan^{-1}{(d t)}\,,\quad\quad 0\leq t<\infty\, ,
\end{equation}
where $c$ and $d$ are, again, some real parameters. We refer to this contour as the \textit{Self Energy Contour (SEC)}.
\begin{figure}[t]
    \includegraphics[height=5.1cm]{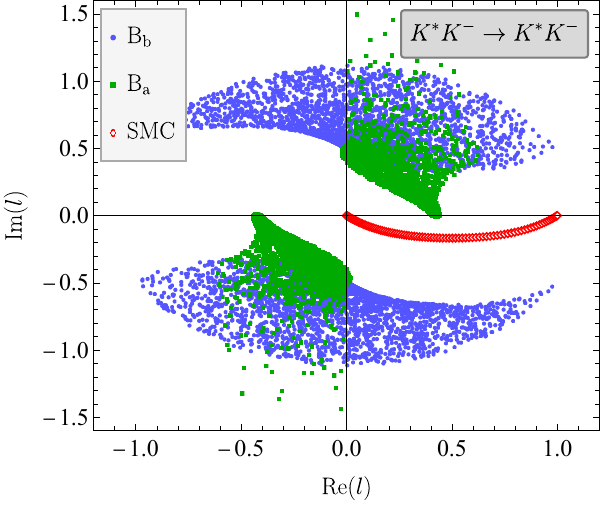}
    \includegraphics[height=5.1cm,trim=1cm 0 0 0,clip]{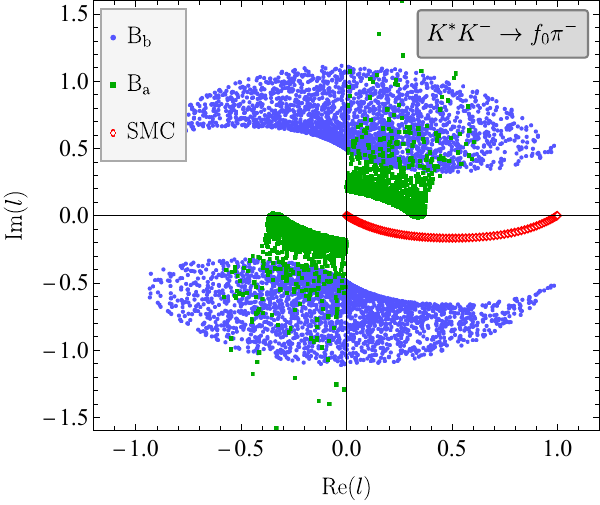}
    \includegraphics[height=5.1cm,trim=1cm 0 0 0,clip]{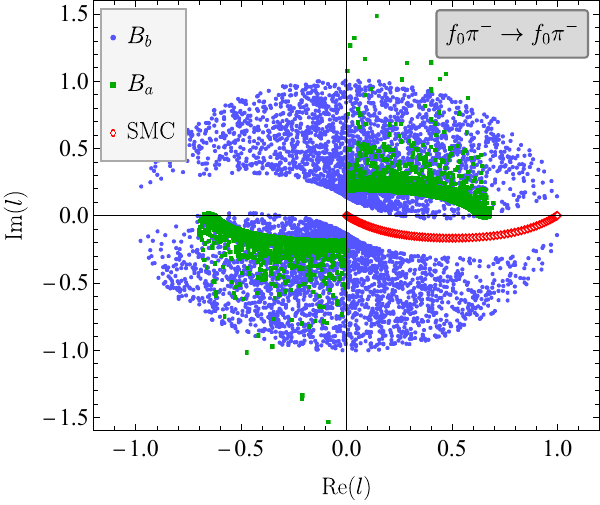}
    \\
    \vspace{4mm}
    \includegraphics[height=5.2cm]{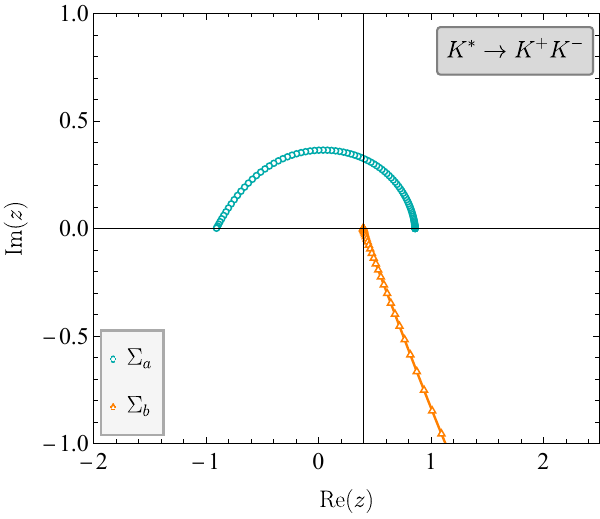}
    \includegraphics[height=5.2cm,trim=1cm 0 0 0,clip]{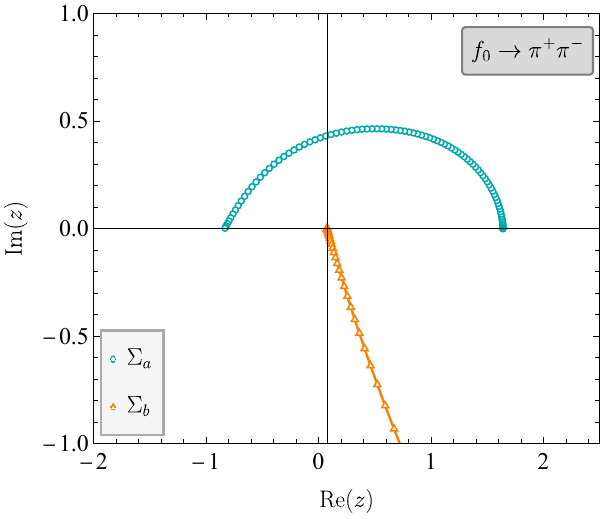}
    \includegraphics[height=5.2cm,trim=1cm 0 0 0,clip]{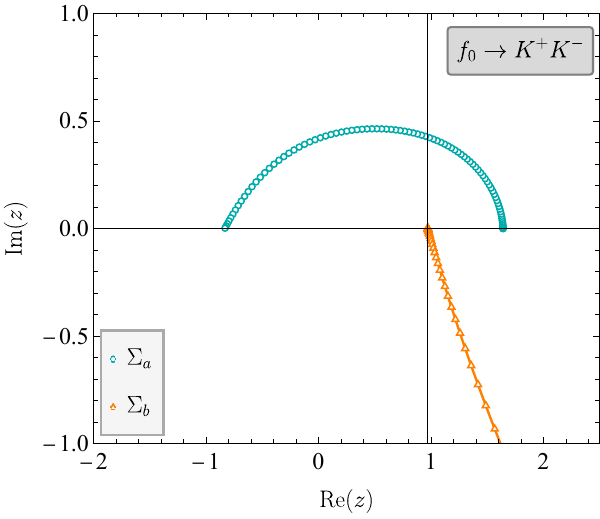}
    \caption{In the top row, the singularities of the exchange term $B$ with respect to the integration momentum $l$ are shown, with the other momentum $p$ fixed on the SMC and $-1\leq\cos\theta\leq 1$. The first, second and third columns correspond to the matrix elements $B_{12}$, $B_{21}$ and $B_{22}$ in the channel space. In the bottom row, the quantities $\sigma(l)$ and $(\sqrt{k^2 + m_1^2}+\sqrt{k^2 + m_2^2})^2$ (on a generic complex $z$-plane) are shown. The first, second and third rows correspond to the $K\pi$, $K\bar K$ and $\pi\pi$ loops.}
    \label{fig:zeros1}  
\end{figure}
In the top row of \cref{fig:zeros1}, it is verified that our choice for the SMC avoids all singularities in the integral. To this end, we define $B_a = \{l \mid E_{\bm{l}+\bm{p}} = 0 \}$ and $B_b = \{l \mid \sqrt{s} - E_{\bm{l}} - E_{\bm{p}} - E_{\bm{l}+\bm{p}} = 0 \}$, see \cref{eqn:b-term}. \cref{fig:zeros1} is then obtained by fixing $p$ along SMC and varying $-1\leq \cos\theta\leq 1$, which are given by the green and blue markers, respectively. It is now clear that the contour, given by red markers, does not overlap with the singularities.

Furthermore, it can be seen from \cref{eqn:tau-se} that the singularities in the self-energy integral emerge exactly for those values of $k$, where $\sigma(l)$ equals to $(\sqrt{k^2 + m_1^2}+\sqrt{k^2 + m_2^2})^2$. To this end, we define $\Sigma_a = \{\sigma(l) \mid l \in \textrm{SMC} \}$ and $\Sigma_b = \{(\sqrt{k^2 + m_1^2}+\sqrt{k^2 + m_2^2})^2 \mid k \in \textrm{SEC} \}$. These quantities are given by the cyan and orange markers in the bottom row of \cref{fig:zeros1}, respectively. It can be seen that these curves do not intersect and, hence, the denominator is not singular for our choice of SMC and SEC.
\item \emph{Discretize the integration interval and solve the integral equation}. The latter step boils down to solving the matrix equation
\begin{equation}
    T = B(\mathbb{1}-\tau B)^{-1}\,
    \label{eq:Fredholm/resolvent}
\end{equation}
for complex spectator momenta on the SMC.
\item \emph{Carry out the integral in \cref{eqn:connected-T} to obtain $\Gamma(\sqrt{s},q \in \mathbb{C})$.} For momenta on the SMC, this amounts to evaluating the matrix equation
\begin{equation}
    \Gamma=T\tau D\,,
    \label{eq:Gamma-operator}
\end{equation}
where $\Gamma$ is a vector of the dimension equal to the number of discrete points chosen in step 2.
\item \emph{Determine $\Gamma$ for real momenta.} This is an unavoidable step when comparing with experimentally accessible line-shapes. Various methods exist to achieve this goal~\cite{SchmidZiegelmann, Hetherington:1965zza, Cahill:1971ddy, Adhikari:1974fh, Matsuyama:2006rp}. We will concentrate on two of them which we find the most convenient: the Rational Analytic Continuation (RAC) method (\cref{sec:RAC-method}) and the Cahill~\&~Sloan (CS) method (\cref{sec:CS-method}). For a discussion and comparison of these and other methods, see also \cite{Feng:2024wyg}.
\end{enumerate}

\subsubsection{Rational Analytic Continuation (RAC) Method}
\label{sec:RAC-method}
A very transparent and flexible method to obtain the decay amplitudes for real momenta relies on the analyticity of the decay amplitudes in complex momenta. Specifically, the obtained result in the complex momentum-plane is analytically continued to the real axis using an analytic function. Typically, for the latter, Padé-approximants are chosen with coefficients fit to the values $\Gamma(q\in SMC)$, and then evaluated for real momenta. Obviously, one needs to ensure that no singularities are located in the extrapolation region (${\rm SMC} \to \mathds{R}$). For recent applications see, e.g., Ref.~\cite{Sadasivan:2020syi, Sadasivan:2021emk}.

In our case, we found that even though Padé-approximants lead to reasonable results, it lacked the ability to reproduce sharp structures, like cusps, in the amplitude. 
We found that (generalized) continued fractions are better at reproducing the above-mentioned singularities. For relations between Padé-approximants and continued fractions, see Ref.~\cite{ANGELL2010904,LORENTZEN20101364}. Here, we make use of Thiele's interpolation formula, which for amplitudes evaluated at complex outgoing spectator momenta $z_1, \dots, z_n$, takes the following form~\cite{Abramowitz-Stegun}:
\begin{align}
    \label{eqn:cfrac}
    T(z)=\rho_0(z_1)
    +\cfrac{z-z_1}{\rho_1(z_1,z_2)+\cfrac{z-z_2}
    {\rho_2(z_1,z_2,z_3) - \rho_0(z_1)+\cfrac{z-z_3}
    {\rho_3(z_1,z_2,z_3,z_4)-\rho_1(z_1,z_2)+\cdots}}}\,,
\end{align}
where the reciprocal differences denoted by $\rho$ are given by
\begin{equation}
\begin{aligned}
    \rho_0(z_1) &= T(z_1),\\
    \rho_1(z_1,z_2) &= \frac{z_1 - z_2}{\rho_0(z_1) - \rho_0(z_2)},\\
    \rho_2(z_1,z_2,z_3) &=  \frac{z_1 - z_3}{\rho_1(z_1,z_2) - \rho_1(z_2,z_3)} + \rho_0(z_2),\\
    &\ \vdots \\
    \rho_n(z_1,z_2,\cdots,z_{n+1})&=\frac{z_1-z_{n+1}}{\rho_{n-1}(z_1,z_2,\cdots,z_n)-\rho_{n-1}(z_2,z_3,\cdots,z_{n+1})} +  \rho_{n-2}(z_2,z_3,\cdots,z_n)\,.
\end{aligned}
\end{equation}
Note that the last term of the continued fraction contains the reciprocal difference $\rho_{n-1}$, in our case. Now, the amplitude given by \cref{eqn:cfrac} is, in principle, valid for all complex values of outgoing spectator momentum $q$, including real values. This reveals another technical advantage of using continued fraction, namely that no interim fit is required, but rather an exact form of the constants $\rho_n$ is available. This improves the performance and allows one to explore systematic uncertainties by increasing, for example, the number of discretization points.

\subsubsection{Cahill \& Sloan (CS) Method}
\label{sec:CS-method}
\begin{figure}[t]
    \centering
    \includegraphics[height=7.5cm]{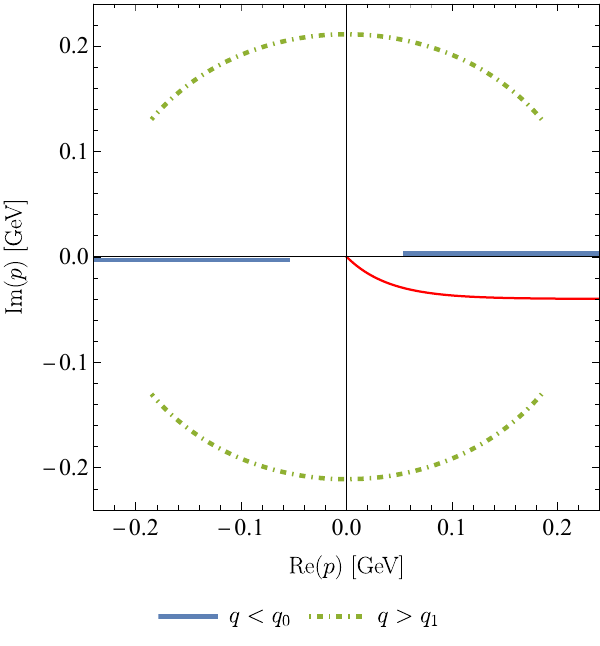}
    \hfil
    \includegraphics[height=7.5cm]{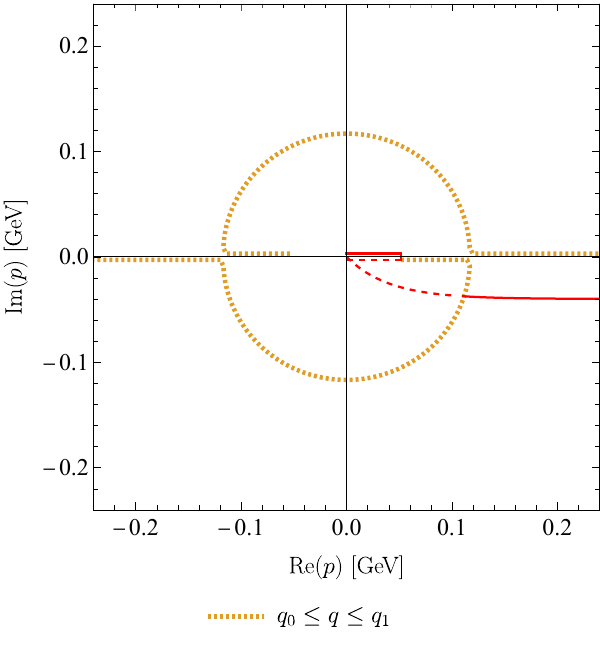}
    \caption{For $q<q_0$ for $q>q_1$, the integral given in \cref{eqn:connected-T} can be evaluated along a simple deformed contour shown on left panel. For $q_0<q<q_1$, one has to travel along the contour shown in the right panel that lies, in part, on the second Riemann sheet (indicated by the dashed line).}
    \label{fig:m1-regions}
\end{figure}

This method was first introduced in Ref.~\cite{Cahill:1971ddy}. One begins with the quantity $\Gamma(\sqrt{s},q)$ again, given by \cref{eqn:connected-T}, for values $q$ on the SMC, which again avoids the singularities as described before. Then one uses the integral representation given by \cref{eqn:connected-T} and analytically continues to the real axis by using Cauchy's theorem. Following the discussion in \cref{sec:IVU-Formalism}, three different scenarios are possible, depending on the numerical value of the momentum $q$ after the continuation, see \cref{fig:branch-cuts}. Namely, if $q<q_0$, where $q_0$ is a critical value expressed through the masses and external kinematic variables, then the kernel has linear cuts shown by the solid blue lines on the left panel of this figure. The analytic continuation is then straightforward --- one merely substitutes real value of $q$ in this integral representation. When $q\geq q_0$, circular cuts are formed that touch the real axis (orange dots on the same figure). This continues until $q\leq q_1$, where $q_1$ is another critical value. After which, 
the circular cuts are pushed away and do not touch the real axis anymore. In this case, again, the analytic continuation is trivial --- one merely assumes $q$ to be real in the integral representation again. The situation for $q<q_0$ and $q>q_1$ is shown in the left panel of \cref{fig:m1-regions}. It is seen that the integration along the original SMC contour is allowed, since it never crosses the singularities. Also note that, for any particular choice of SMC, the contour does not cross the branch cut after $q>q_1+\delta$, where $\delta$ is finite. This quantity can be made arbitrarily small, choosing SMC very close to the real axis. For a detailed discussion, see Ref.~\cite{Pang:2023jri,Feng:2024wyg}.

Thus, only the region $q_0<q<q_1$ is problematic. The original integration contour has to be deformed, avoiding the singularities of the kernel, and also accommodate the different Riemann sheets corresponding to the logarithmic branch cut arising from \cref{eqn:b-integrated}. The deformed contour is shown in the right panel of \cref{fig:m1-regions}. It goes from the origin up to the branch point, after which one moves to the second sheet and goes back to origin. This is followed by moving along the original SMC up to the branch cut, after which one moves to the first sheet again. Note that, in order to evaluate this integral, one needs the values of the integrand on the real axis for values below the branch point. Still, these can be easily calculated, since the analytic continuation in this region does not encounter any singularity. More information about this method can be found in the original paper~\cite{Cahill:1971ddy} as well as in the textbook~\cite{SchmidZiegelmann}. For a recent application and further details, see Refs.~\cite{Pang:2023jri,Feng:2024wyg}.

We note one drawback of the CS method. Namely, when there are two or more channels, there might be an overlap of the singularity region of a particular channel with the small $3$-momenta region of another channel. Problems arise, when the calculations are done for the $3$-momenta values in the overlapping regions. Fortunately, this is not the case here in the region of energies we are interested in. Still, in general, this method is not well suited for systems with multiple channels and certain care is needed to implement it in this case. For further discussions of different methods, we refer the interested reader again to the recent work~\cite{Feng:2024wyg}.

\begin{figure}[t]
    \centering
    \includegraphics[width=0.42\linewidth]{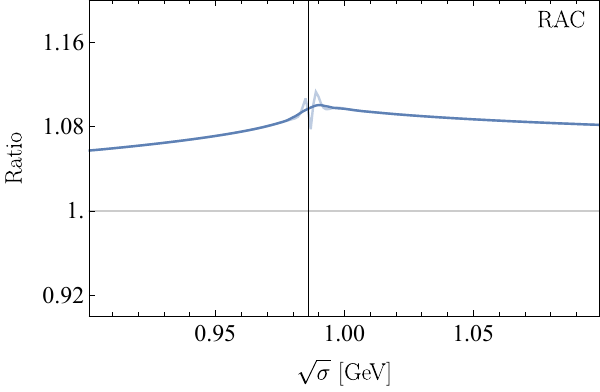}
    \hfil
    \includegraphics[width=0.42\linewidth]{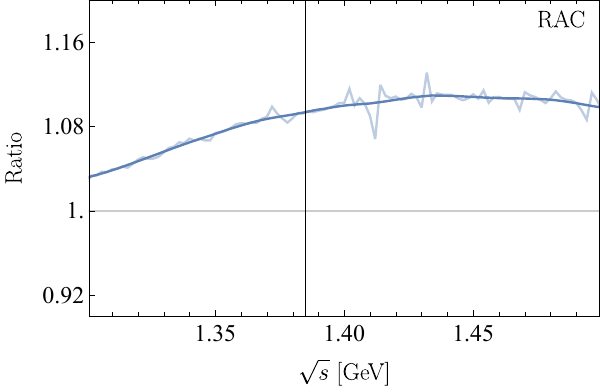}
    \\
    \vspace{4mm}
    \includegraphics[width=0.42\linewidth]{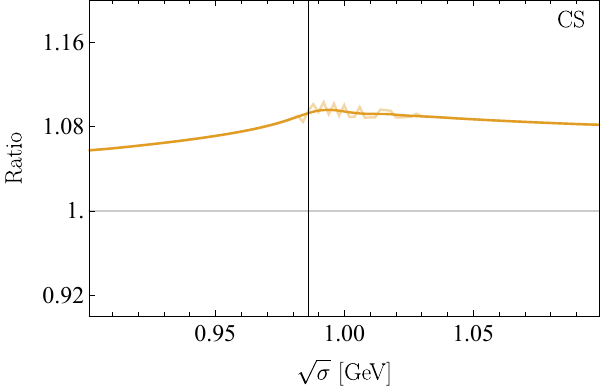}
    \hfil
    \includegraphics[width=0.42\linewidth]{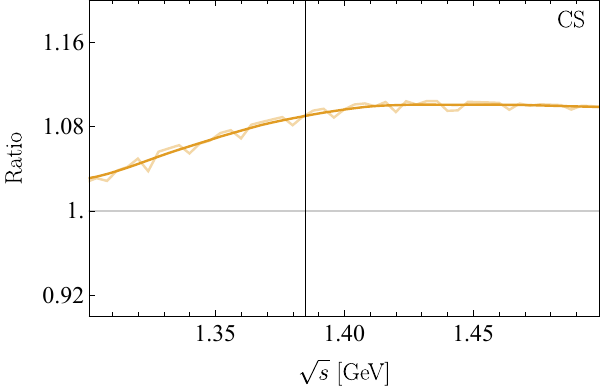}
    \caption{The numerical oscillations in the full solution in RAC and CS methods. We plot the ratio $|\Gamma/\Gamma_\triangle|^2$, both the exact and the smoothed versions. It is seen that the oscillations are indeed small and do not affect the conclusions.}
    \label{fig:comparison}
\end{figure}

Additionally, we would like to emphasize that both the presented methods boil down to numerical evaluations. Thus, some instabilities due to finite resolution are to be expected. Interestingly, we found that both methods show similar performance in this regard, see \cref{fig:comparison}, where we show the ratio of the ``naked'' ($\Gamma_\triangle $) and ``fully dressed'' ($\Gamma$) amplitudes calculated with both methods. As the calculations become quite time-consuming for a large number of mesh points, we have opted to display the ``smoothed'' results everywhere in the following to make the interpretation of the figures easier. The smoothed results are obtained by convolving the obtained final amplitudes with a Gaussian kernel of varying standard deviations of $2$ to $5$.

\subsection{Results}
\label{subsec:results}
\begin{figure}
    \centering
    \includegraphics[width=0.42\linewidth]{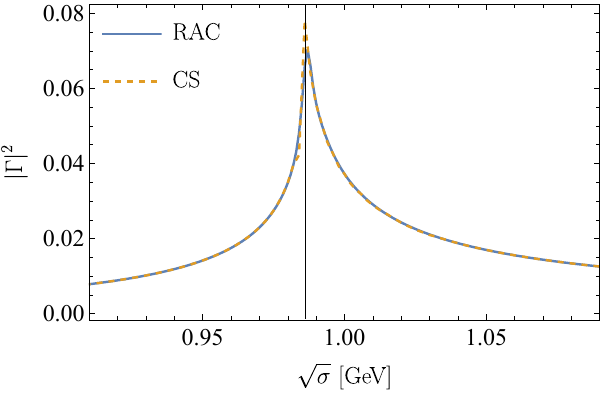}
    \hfil
    \includegraphics[width=0.42\linewidth]{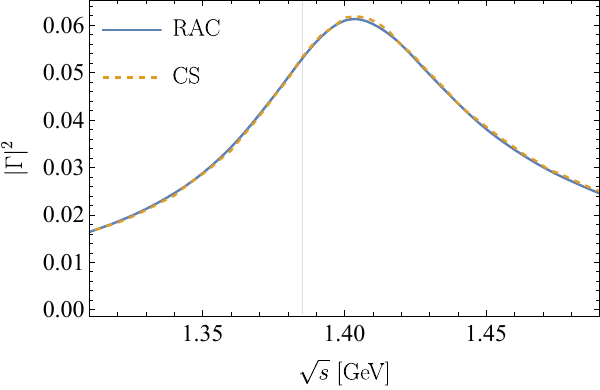}
    \caption{On the left, comparison of the decay amplitudes squared as a function of the outgoing isobar invariant mass $\sqrt{\sigma}$ for a fixed incoming invariant mass $\sqrt{s} = 1.42~\textrm{GeV}$ between RAC and CS methods. On the right, the comparison for a fixed $\sqrt{\sigma} = 0.99~\textrm{GeV}$ and varying $\sqrt{s}$.}
    \label{fig:m1-compare}
\end{figure}

We now discuss the main results of this work, namely the study of final-state interactions in relation to the triangle singularity. The result including all final-state interactions is shown in \cref{fig:m1-compare}, which shows the full decay amplitudes squared $|\Gamma(\sqrt{s},\bm{q})|^2$ as a function of the variables $s$ and $\sigma$. There, both the methods (RAC and CS) are very similar to each other, except the fact that the RAC method is slightly smeared out in the immediate vicinity of singularities.
Overall, it is safe to say that the triangle singularity persists, as expected, when all final-state interactions are taken into account. At first glance, it might seem surprising that there is no unitary cusp in the variable $s$ (the right panel of \cref{fig:m1-compare}). A close scrutiny of the problem reveals the reason for it: the unitary cusp is completely overshadowed by the huge triangle singularity, with both effects being smeared out by the finite width of the $K^*$.

\begin{figure}[!t]
    \includegraphics[width=0.42\linewidth]{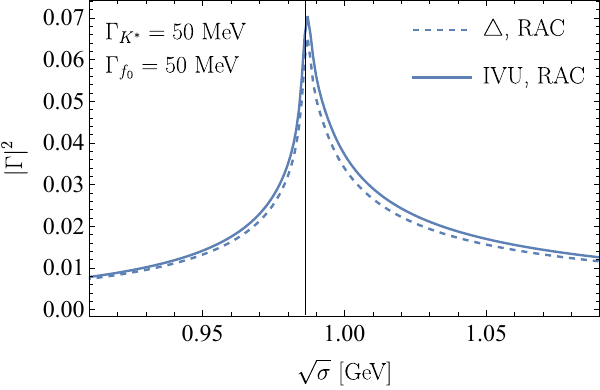}
    \hfil
    \includegraphics[width=0.42\linewidth]{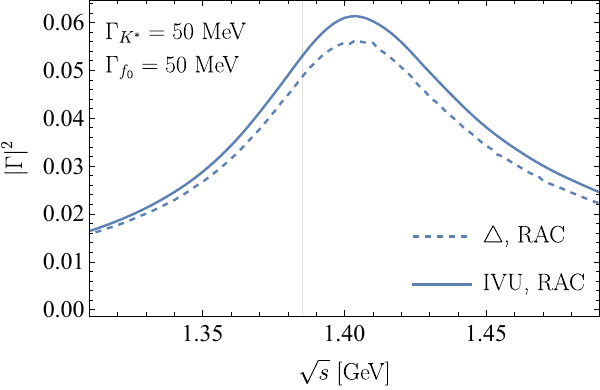}
    \caption{On the left, the decay amplitudes squared for the triangle diagram as a function of the outgoing isobar invariant mass $\sqrt{\sigma}$ for a fixed incoming invariant mass $\sqrt{s}=1.42~\textrm{GeV}$. On the right, the same quantity for $\sqrt{\sigma}=0.99~\textrm{GeV}$ and varying $\sqrt{s}$. The plots show the amplitudes for the triangle and the fully dressed diagrams. For simplicity, we restrict ourselves to the RAC method.}
    \label{fig:triangle-ivu}
\end{figure}

\begin{figure}[!t]
    \centering
    \includegraphics[height=4.8cm,trim=0 0 5cm 0,clip]{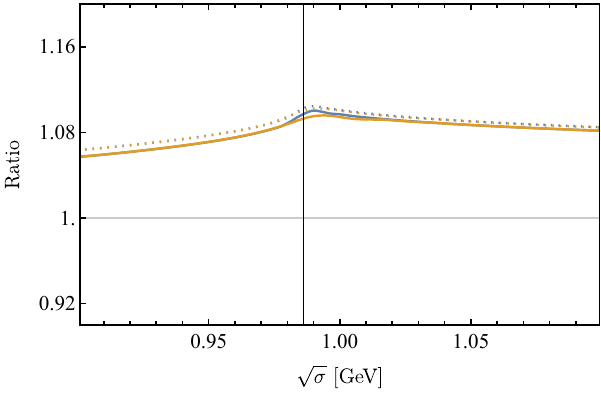}
    \hfil
    \includegraphics[height=4.8cm,trim=0.65cm 0 0 0,clip]{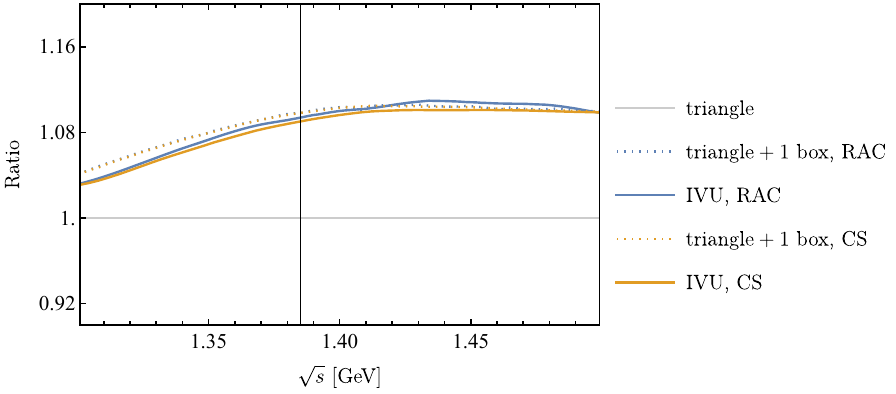}
    \caption{The ratio  $|\Gamma/\Gamma_\triangle|^2$ for the first Born approximation (triangle plus one box) and the full solution. It is seen that the results in both the RAC and CS methods are very similar, and the first approximation is almost the same as the full solution.}
    \label{fig:comparison1}
\end{figure}

To quantify the effect of final-state interactions, we plot in \cref{fig:triangle-ivu} the ``naked'' ($|\Gamma_\triangle|^2$) and ``fully dressed'' ($|\Gamma|^2$) full decay amplitudes squared as functions of incoming invariant mass $\sqrt{s}$ for a fixed outgoing isobar invariant mass $\sqrt{\sigma}$, and vice versa. We observe that the rescattering corrections are rather small. This suggests that the Born series~\cref{eq:series} is converging fast, and justifies truncating the series at the leading order. More generally, the rapidity of the convergence of the Born series can be seen in \cref{fig:comparison1}. There we plot the full decay amplitude squared of the $a_1$ as well as the triangle diagram plus one box normalized by the ``naked'' amplitude, i.e., the triangle diagram without final-state interactions. It shows that (a) the corrections to the decay amplitude are of the order of 10\% (in our toy model), and (b) the Born series is converging very fast, the first Born approximation almost coinciding with the final result.
    
At this point, we mention that a preliminary study of the same system with particles carrying spin and isospin has been carried out, using the methods from Ref.~\cite{Sadasivan:2020syi, Sadasivan:2021emk}. The results of these studies confirm the results obtained in our toy model. The suppression of the rescattering effects is still present, albeit not so pronounced as before.

\section{Conclusions}
\label{sec:conclusions}
In the present paper, we have studied the effect of the infinite two- and three-body rescattering on the triangle singularity. Such singularity arises for certain configurations of masses of involved particles and in some specific kinematic conditions. The formalism used for this study was completely general. We chose to work with the toy model that mimicked the essential features of the decay of the would-be $a_1(1420)$ meson, see Ref.~\cite{Mikhasenko:2015oxp}.

In the beginning, using Landau equations, we have shown that the triangle singularity on the real axis also arises in the multiloop ladder diagrams that describe final-state interactions. This can be understood by considering multiple contractions of internal propagators in the ladder diagrams attached to the triangle. Either such contractions lead to non-singular configurations or provide a subleading singularity through a factorization into the triangle diagram and the rest that is non-singular. Thus, the singularity arises exactly at the same place as in the simple triangle diagram.

The analysis through Landau equations is quite general. Still, it provides one only with the location of the singularity and does not give access to its strength or corresponding line-shapes. In the second part of the paper, we have carried out a full-fledged analysis of a problem involving unstable particles (isobars). To this end, the unitary three-body formalism (IVU)~\cite{Mai:2017vot} has been used, which was previously applied to the analysis of the $a_1(1260)$ in Refs.~\cite{Sadasivan:2020syi,Sadasivan:2021emk,Mai:2021nul}. At the first step, we have ignored the spin and isospin of all particles involved, albeit extended our toy model to the multichannel case that is essential to allow the emergence of the triangle singularity. Within the approximations made, we have found that the Born series converges rather rapidly, leaving final-state interactions as sub-dominant to the triangle diagram. We emphasize again that our aim here was not to give an exact quantitative calculation of the effect of final-state rescattering in this particular physical system, but rather to set and to test a framework for such calculations that are planned in the near future. Still, one might expect that the results obtained within the toy model qualitatively describe the gross picture that emerges, if all neglected effects are taken into account.

Last but not least, let us note that the development of a quantitative framework which describes the system in question allows one, eventually, to reformulate the framework in a finite volume. This, in its turn, could be useful for the analysis of lattice data in the channel with quantum numbers of the $a_1$, extending the energy region beyond $a_1(1260)$ which was studied already in Ref.~\cite{Mai:2021nul}.

\bigskip

\begin{acknowledgments}
We thank Melahat~Bayar, Rishabh~Bubna, Feng-Kun~Guo, Bernhard~Ketzer, Matthias~Lutz, Eric~Swanson, Mathias~Wagner and Qiang~Zhao for useful discussions and motivation to pursue the topic of this work. 
MM and AR acknowledge funding by the Deutsche Forschungsgemeinschaft (DFG, German Research Foundation) through the Sino-German Collaborative Research Center TRR110 “Symmetries and the Emergence of Structure in QCD” (DFG Project ID 196253076 - TRR 110). MM acknowledges funding through the Heisenberg Programme (project number: 532635001).
The work of MD and MM is supported by the National Science Foundation (NSF) Grant No. 2310036. MD is also supported by the U.S. Department of Energy grant DE-SC0016582, and Office of Science, Office of Nuclear Physics under contract DE-AC05- 06OR23177. This work contributes to the aims of the U.S. Department of Energy ExoHad Topical Collaboration, contract DE-SC0023598. 
The work AR and AS was funded in part by the Ministry of Culture and Science of North Rhine-Westphalia through the NRW-FAIR project. A.R., in addition, thanks the Chinese Academy of Sciences (CAS) President’s International Fellowship Initiative (PIFI) (grant no. 2024VMB0001) for the partial financial support.
\end{acknowledgments}
\footnotesize
\bibliography{BIB,NON-INSPIRE}

\begin{thebibliography}{90}%
\makeatletter
\providecommand \@ifxundefined [1]{%
 \@ifx{#1\undefined}
}%
\providecommand \@ifnum [1]{%
 \ifnum #1\expandafter \@firstoftwo
 \else \expandafter \@secondoftwo
 \fi
}%
\providecommand \@ifx [1]{%
 \ifx #1\expandafter \@firstoftwo
 \else \expandafter \@secondoftwo
 \fi
}%
\providecommand \natexlab [1]{#1}%
\providecommand \enquote  [1]{``#1''}%
\providecommand \bibnamefont  [1]{#1}%
\providecommand \bibfnamefont [1]{#1}%
\providecommand \citenamefont [1]{#1}%
\providecommand \href@noop [0]{\@secondoftwo}%
\providecommand \href [0]{\begingroup \@sanitize@url \@href}%
\providecommand \@href[1]{\@@startlink{#1}\@@href}%
\providecommand \@@href[1]{\endgroup#1\@@endlink}%
\providecommand \@sanitize@url [0]{\catcode `\\12\catcode `\$12\catcode `\&12\catcode `\#12\catcode `\^12\catcode `\_12\catcode `\%12\relax}%
\providecommand \@@startlink[1]{}%
\providecommand \@@endlink[0]{}%
\providecommand \url  [0]{\begingroup\@sanitize@url \@url }%
\providecommand \@url [1]{\endgroup\@href {#1}{\urlprefix }}%
\providecommand \urlprefix  [0]{URL }%
\providecommand \Eprint [0]{\href }%
\providecommand \doibase [0]{http://dx.doi.org/}%
\providecommand \selectlanguage [0]{\@gobble}%
\providecommand \bibinfo  [0]{\@secondoftwo}%
\providecommand \bibfield  [0]{\@secondoftwo}%
\providecommand \translation [1]{[#1]}%
\providecommand \BibitemOpen [0]{}%
\providecommand \bibitemStop [0]{}%
\providecommand \bibitemNoStop [0]{.\EOS\space}%
\providecommand \EOS [0]{\spacefactor3000\relax}%
\providecommand \BibitemShut  [1]{\csname bibitem#1\endcsname}%
\let\auto@bib@innerbib\@empty
\bibitem [{\citenamefont {Mai}\ \emph {et~al.}(2023)\citenamefont {Mai}, \citenamefont {Mei\ss{}ner},\ and\ \citenamefont {Urbach}}]{Mai:2022eur}%
  \BibitemOpen
  \bibfield  {author} {\bibinfo {author} {\bibfnamefont {Maxim}\ \bibnamefont {Mai}}, \bibinfo {author} {\bibfnamefont {Ulf-G.}\ \bibnamefont {Mei\ss{}ner}}, \ and\ \bibinfo {author} {\bibfnamefont {Carsten}\ \bibnamefont {Urbach}},\ }\bibfield  {title} {\enquote {\bibinfo {title} {{Towards a theory of hadron resonances}},}\ }\href {\doibase 10.1016/j.physrep.2022.11.005} {\bibfield  {journal} {\bibinfo  {journal} {Phys. Rept.}\ }\textbf {\bibinfo {volume} {1001}},\ \bibinfo {pages} {1--66} (\bibinfo {year} {2023})},\ \Eprint {http://arxiv.org/abs/2206.01477} {arXiv:2206.01477 [hep-ph]} \BibitemShut {NoStop}%
\bibitem [{\citenamefont {Lutz}\ \emph {et~al.}(2015)\citenamefont {Lutz}, \citenamefont {Kolomeitsev},\ and\ \citenamefont {Korpa}}]{Lutz:2015lca}%
  \BibitemOpen
  \bibfield  {author} {\bibinfo {author} {\bibfnamefont {M.~F.~M.}\ \bibnamefont {Lutz}}, \bibinfo {author} {\bibfnamefont {E.~E.}\ \bibnamefont {Kolomeitsev}}, \ and\ \bibinfo {author} {\bibfnamefont {C.~L.}\ \bibnamefont {Korpa}},\ }\bibfield  {title} {\enquote {\bibinfo {title} {{Spectral representation for $u$- and $t$-channel exchange processes in a partial-wave decomposition}},}\ }\href {\doibase 10.1103/PhysRevD.92.016003} {\bibfield  {journal} {\bibinfo  {journal} {Phys. Rev. D}\ }\textbf {\bibinfo {volume} {92}},\ \bibinfo {pages} {016003} (\bibinfo {year} {2015})},\ \Eprint {http://arxiv.org/abs/1506.02375} {arXiv:1506.02375 [hep-ph]} \BibitemShut {NoStop}%
\bibitem [{\citenamefont {Lutz}\ and\ \citenamefont {Korpa}(2018)}]{Lutz:2018kaz}%
  \BibitemOpen
  \bibfield  {author} {\bibinfo {author} {\bibfnamefont {M.~F.~M.}\ \bibnamefont {Lutz}}\ and\ \bibinfo {author} {\bibfnamefont {C.~L.}\ \bibnamefont {Korpa}},\ }\bibfield  {title} {\enquote {\bibinfo {title} {{On coupled-channel dynamics in the presence of anomalous thresholds}},}\ }\href {\doibase 10.1103/PhysRevD.98.076003} {\bibfield  {journal} {\bibinfo  {journal} {Phys. Rev. D}\ }\textbf {\bibinfo {volume} {98}},\ \bibinfo {pages} {076003} (\bibinfo {year} {2018})},\ \Eprint {http://arxiv.org/abs/1808.08695} {arXiv:1808.08695 [hep-ph]} \BibitemShut {NoStop}%
\bibitem [{\citenamefont {Korpa}\ \emph {et~al.}(2023)\citenamefont {Korpa}, \citenamefont {Lutz}, \citenamefont {Guo},\ and\ \citenamefont {Heo}}]{Korpa:2022voo}%
  \BibitemOpen
  \bibfield  {author} {\bibinfo {author} {\bibfnamefont {Csaba~L.}\ \bibnamefont {Korpa}}, \bibinfo {author} {\bibfnamefont {Matthias F.~M.}\ \bibnamefont {Lutz}}, \bibinfo {author} {\bibfnamefont {Xiao-Yu}\ \bibnamefont {Guo}}, \ and\ \bibinfo {author} {\bibfnamefont {Yonggoo}\ \bibnamefont {Heo}},\ }\bibfield  {title} {\enquote {\bibinfo {title} {{Coupled-channel system with anomalous thresholds and unitarity}},}\ }\href {\doibase 10.1103/PhysRevD.107.L031505} {\bibfield  {journal} {\bibinfo  {journal} {Phys. Rev. D}\ }\textbf {\bibinfo {volume} {107}},\ \bibinfo {pages} {L031505} (\bibinfo {year} {2023})},\ \Eprint {http://arxiv.org/abs/2211.03508} {arXiv:2211.03508 [hep-ph]} \BibitemShut {NoStop}%
\bibitem [{\citenamefont {Richard}(2012)}]{Richard:2012xw}%
  \BibitemOpen
  \bibfield  {author} {\bibinfo {author} {\bibfnamefont {Jean-Marc}\ \bibnamefont {Richard}},\ }\bibfield  {title} {\enquote {\bibinfo {title} {{An introduction to the quark model}},}\ }in\ \href@noop {} {\emph {\bibinfo {booktitle} {{Ferrara International School Niccol\`o Cabeo 2012: Hadronic spectroscopy}}}}\ (\bibinfo {year} {2012})\ \Eprint {http://arxiv.org/abs/1205.4326} {arXiv:1205.4326 [hep-ph]} \BibitemShut {NoStop}%
\bibitem [{\citenamefont {Eichmann}\ \emph {et~al.}(2016{\natexlab{a}})\citenamefont {Eichmann}, \citenamefont {Sanchis-Alepuz}, \citenamefont {Williams}, \citenamefont {Alkofer},\ and\ \citenamefont {Fischer}}]{Eichmann:2016yit}%
  \BibitemOpen
  \bibfield  {author} {\bibinfo {author} {\bibfnamefont {Gernot}\ \bibnamefont {Eichmann}}, \bibinfo {author} {\bibfnamefont {Helios}\ \bibnamefont {Sanchis-Alepuz}}, \bibinfo {author} {\bibfnamefont {Richard}\ \bibnamefont {Williams}}, \bibinfo {author} {\bibfnamefont {Reinhard}\ \bibnamefont {Alkofer}}, \ and\ \bibinfo {author} {\bibfnamefont {Christian~S.}\ \bibnamefont {Fischer}},\ }\bibfield  {title} {\enquote {\bibinfo {title} {{Baryons as relativistic three-quark bound states}},}\ }\href {\doibase 10.1016/j.ppnp.2016.07.001} {\bibfield  {journal} {\bibinfo  {journal} {Prog. Part. Nucl. Phys.}\ }\textbf {\bibinfo {volume} {91}},\ \bibinfo {pages} {1--100} (\bibinfo {year} {2016}{\natexlab{a}})},\ \Eprint {http://arxiv.org/abs/1606.09602} {arXiv:1606.09602 [hep-ph]} \BibitemShut {NoStop}%
\bibitem [{\citenamefont {Eichmann}\ \emph {et~al.}(2016{\natexlab{b}})\citenamefont {Eichmann}, \citenamefont {Fischer},\ and\ \citenamefont {Sanchis-Alepuz}}]{Eichmann:2016hgl}%
  \BibitemOpen
  \bibfield  {author} {\bibinfo {author} {\bibfnamefont {Gernot}\ \bibnamefont {Eichmann}}, \bibinfo {author} {\bibfnamefont {Christian~S.}\ \bibnamefont {Fischer}}, \ and\ \bibinfo {author} {\bibfnamefont {Helios}\ \bibnamefont {Sanchis-Alepuz}},\ }\bibfield  {title} {\enquote {\bibinfo {title} {{Light baryons and their excitations}},}\ }\href {\doibase 10.1103/PhysRevD.94.094033} {\bibfield  {journal} {\bibinfo  {journal} {Phys. Rev. D}\ }\textbf {\bibinfo {volume} {94}},\ \bibinfo {pages} {094033} (\bibinfo {year} {2016}{\natexlab{b}})},\ \Eprint {http://arxiv.org/abs/1607.05748} {arXiv:1607.05748 [hep-ph]} \BibitemShut {NoStop}%
\bibitem [{\citenamefont {Yin}\ \emph {et~al.}(2019)\citenamefont {Yin}, \citenamefont {Chen}, \citenamefont {Krein}, \citenamefont {Roberts}, \citenamefont {Segovia},\ and\ \citenamefont {Xu}}]{Yin:2019bxe}%
  \BibitemOpen
  \bibfield  {author} {\bibinfo {author} {\bibfnamefont {Pei-Lin}\ \bibnamefont {Yin}}, \bibinfo {author} {\bibfnamefont {Chen}\ \bibnamefont {Chen}}, \bibinfo {author} {\bibfnamefont {Gast\~ao}\ \bibnamefont {Krein}}, \bibinfo {author} {\bibfnamefont {Craig~D.}\ \bibnamefont {Roberts}}, \bibinfo {author} {\bibfnamefont {Jorge}\ \bibnamefont {Segovia}}, \ and\ \bibinfo {author} {\bibfnamefont {Shu-Sheng}\ \bibnamefont {Xu}},\ }\bibfield  {title} {\enquote {\bibinfo {title} {{Masses of ground-state mesons and baryons, including those with heavy quarks}},}\ }\href {\doibase 10.1103/PhysRevD.100.034008} {\bibfield  {journal} {\bibinfo  {journal} {Phys. Rev. D}\ }\textbf {\bibinfo {volume} {100}},\ \bibinfo {pages} {034008} (\bibinfo {year} {2019})},\ \Eprint {http://arxiv.org/abs/1903.00160} {arXiv:1903.00160 [nucl-th]} \BibitemShut {NoStop}%
\bibitem [{\citenamefont {Maiani}\ \emph {et~al.}(2005)\citenamefont {Maiani}, \citenamefont {Piccinini}, \citenamefont {Polosa},\ and\ \citenamefont {Riquer}}]{Maiani:2004vq}%
  \BibitemOpen
  \bibfield  {author} {\bibinfo {author} {\bibfnamefont {L.}~\bibnamefont {Maiani}}, \bibinfo {author} {\bibfnamefont {F.}~\bibnamefont {Piccinini}}, \bibinfo {author} {\bibfnamefont {A.~D.}\ \bibnamefont {Polosa}}, \ and\ \bibinfo {author} {\bibfnamefont {V.}~\bibnamefont {Riquer}},\ }\bibfield  {title} {\enquote {\bibinfo {title} {{Diquark-antidiquarks with hidden or open charm and the nature of $X(3872)$}},}\ }\href {\doibase 10.1103/PhysRevD.71.014028} {\bibfield  {journal} {\bibinfo  {journal} {Phys. Rev. D}\ }\textbf {\bibinfo {volume} {71}},\ \bibinfo {pages} {014028} (\bibinfo {year} {2005})},\ \Eprint {http://arxiv.org/abs/hep-ph/0412098} {arXiv:hep-ph/0412098} \BibitemShut {NoStop}%
\bibitem [{\citenamefont {Heupel}\ \emph {et~al.}(2012)\citenamefont {Heupel}, \citenamefont {Eichmann},\ and\ \citenamefont {Fischer}}]{Heupel:2012ua}%
  \BibitemOpen
  \bibfield  {author} {\bibinfo {author} {\bibfnamefont {Walter}\ \bibnamefont {Heupel}}, \bibinfo {author} {\bibfnamefont {Gernot}\ \bibnamefont {Eichmann}}, \ and\ \bibinfo {author} {\bibfnamefont {Christian~S.}\ \bibnamefont {Fischer}},\ }\bibfield  {title} {\enquote {\bibinfo {title} {{Tetraquark Bound States in a Bethe-Salpeter Approach}},}\ }\href {\doibase 10.1016/j.physletb.2012.11.009} {\bibfield  {journal} {\bibinfo  {journal} {Phys. Lett. B}\ }\textbf {\bibinfo {volume} {718}},\ \bibinfo {pages} {545--549} (\bibinfo {year} {2012})},\ \Eprint {http://arxiv.org/abs/1206.5129} {arXiv:1206.5129 [hep-ph]} \BibitemShut {NoStop}%
\bibitem [{\citenamefont {Aaij}\ \emph {et~al.}(2015)\citenamefont {Aaij} \emph {et~al.}}]{LHCb:2015yax}%
  \BibitemOpen
  \bibfield  {author} {\bibinfo {author} {\bibfnamefont {Roel}\ \bibnamefont {Aaij}} \emph {et~al.} (\bibinfo {collaboration} {LHCb}),\ }\bibfield  {title} {\enquote {\bibinfo {title} {{Observation of $J/\psi p$ Resonances Consistent with Pentaquark States in $\Lambda_b^0 \to J/\psi K^- p$ Decays}},}\ }\href {\doibase 10.1103/PhysRevLett.115.072001} {\bibfield  {journal} {\bibinfo  {journal} {Phys. Rev. Lett.}\ }\textbf {\bibinfo {volume} {115}},\ \bibinfo {pages} {072001} (\bibinfo {year} {2015})},\ \Eprint {http://arxiv.org/abs/1507.03414} {arXiv:1507.03414 [hep-ex]} \BibitemShut {NoStop}%
\bibitem [{\citenamefont {Bruns}\ \emph {et~al.}(2011)\citenamefont {Bruns}, \citenamefont {Mai},\ and\ \citenamefont {Mei{\ss}ner}}]{Bruns:2010sv}%
  \BibitemOpen
  \bibfield  {author} {\bibinfo {author} {\bibfnamefont {Peter~C.}\ \bibnamefont {Bruns}}, \bibinfo {author} {\bibfnamefont {Maxim}\ \bibnamefont {Mai}}, \ and\ \bibinfo {author} {\bibfnamefont {Ulf-G.}\ \bibnamefont {Mei{\ss}ner}},\ }\bibfield  {title} {\enquote {\bibinfo {title} {{Chiral dynamics of the $S_{11}(1535)$ and $S_{11}(1650)$ resonances revisited}},}\ }\href {\doibase 10.1016/j.physletb.2011.02.008} {\bibfield  {journal} {\bibinfo  {journal} {Phys. Lett. B}\ }\textbf {\bibinfo {volume} {697}},\ \bibinfo {pages} {254--259} (\bibinfo {year} {2011})},\ \Eprint {http://arxiv.org/abs/1012.2233} {arXiv:1012.2233 [nucl-th]} \BibitemShut {NoStop}%
\bibitem [{\citenamefont {Lutz}\ and\ \citenamefont {Kolomeitsev}(2002)}]{Lutz:2001yb}%
  \BibitemOpen
  \bibfield  {author} {\bibinfo {author} {\bibfnamefont {M.~F.~M.}\ \bibnamefont {Lutz}}\ and\ \bibinfo {author} {\bibfnamefont {E.~E.}\ \bibnamefont {Kolomeitsev}},\ }\bibfield  {title} {\enquote {\bibinfo {title} {{Relativistic chiral $SU(3)$ symmetry, large $N_c$ sum rules and meson baryon scattering}},}\ }\href {\doibase 10.1016/S0375-9474(01)01312-4} {\bibfield  {journal} {\bibinfo  {journal} {Nucl. Phys. A}\ }\textbf {\bibinfo {volume} {700}},\ \bibinfo {pages} {193--308} (\bibinfo {year} {2002})},\ \Eprint {http://arxiv.org/abs/nucl-th/0105042} {arXiv:nucl-th/0105042} \BibitemShut {NoStop}%
\bibitem [{\citenamefont {Oset}\ \emph {et~al.}(2016)\citenamefont {Oset} \emph {et~al.}}]{Oset:2016lyh}%
  \BibitemOpen
  \bibfield  {author} {\bibinfo {author} {\bibfnamefont {Eulogio}\ \bibnamefont {Oset}} \emph {et~al.},\ }\bibfield  {title} {\enquote {\bibinfo {title} {{Weak decays of heavy hadrons into dynamically generated resonances}},}\ }\href {\doibase 10.1142/S0218301316300010} {\bibfield  {journal} {\bibinfo  {journal} {Int. J. Mod. Phys. E}\ }\textbf {\bibinfo {volume} {25}},\ \bibinfo {pages} {1630001} (\bibinfo {year} {2016})},\ \Eprint {http://arxiv.org/abs/1601.03972} {arXiv:1601.03972 [hep-ph]} \BibitemShut {NoStop}%
\bibitem [{\citenamefont {Mai}\ and\ \citenamefont {Mei{\ss}ner}(2013)}]{Mai:2012dt}%
  \BibitemOpen
  \bibfield  {author} {\bibinfo {author} {\bibfnamefont {Maxim}\ \bibnamefont {Mai}}\ and\ \bibinfo {author} {\bibfnamefont {Ulf-G.}\ \bibnamefont {Mei{\ss}ner}},\ }\bibfield  {title} {\enquote {\bibinfo {title} {{New insights into antikaon-nucleon scattering and the structure of the $\Lambda(1405)$}},}\ }\href {\doibase 10.1016/j.nuclphysa.2013.01.032} {\bibfield  {journal} {\bibinfo  {journal} {Nucl. Phys. A}\ }\textbf {\bibinfo {volume} {900}},\ \bibinfo {pages} {51 -- 64} (\bibinfo {year} {2013})},\ \Eprint {http://arxiv.org/abs/1202.2030} {arXiv:1202.2030 [nucl-th]} \BibitemShut {NoStop}%
\bibitem [{\citenamefont {Mai}\ and\ \citenamefont {Mei\ss{}ner}(2015)}]{Mai:2014xna}%
  \BibitemOpen
  \bibfield  {author} {\bibinfo {author} {\bibfnamefont {Maxim}\ \bibnamefont {Mai}}\ and\ \bibinfo {author} {\bibfnamefont {Ulf-G.}\ \bibnamefont {Mei\ss{}ner}},\ }\bibfield  {title} {\enquote {\bibinfo {title} {{Constraints on the chiral unitary $\bar KN$ amplitude from $\pi\Sigma K^+$ photoproduction data}},}\ }\href {\doibase 10.1140/epja/i2015-15030-3} {\bibfield  {journal} {\bibinfo  {journal} {Eur. Phys. J. A}\ }\textbf {\bibinfo {volume} {51}},\ \bibinfo {pages} {30} (\bibinfo {year} {2015})},\ \Eprint {http://arxiv.org/abs/1411.7884} {arXiv:1411.7884 [hep-ph]} \BibitemShut {NoStop}%
\bibitem [{\citenamefont {Guo}\ \emph {et~al.}(2015)\citenamefont {Guo}, \citenamefont {Mei\ss{}ner}, \citenamefont {Wang},\ and\ \citenamefont {Yang}}]{Guo:2015umn}%
  \BibitemOpen
  \bibfield  {author} {\bibinfo {author} {\bibfnamefont {Feng-Kun}\ \bibnamefont {Guo}}, \bibinfo {author} {\bibfnamefont {Ulf-G.}\ \bibnamefont {Mei\ss{}ner}}, \bibinfo {author} {\bibfnamefont {Wei}\ \bibnamefont {Wang}}, \ and\ \bibinfo {author} {\bibfnamefont {Zhi}\ \bibnamefont {Yang}},\ }\bibfield  {title} {\enquote {\bibinfo {title} {{How to reveal the exotic nature of the P$_c$(4450)}},}\ }\href {\doibase 10.1103/PhysRevD.92.071502} {\bibfield  {journal} {\bibinfo  {journal} {Phys. Rev. D}\ }\textbf {\bibinfo {volume} {92}},\ \bibinfo {pages} {071502} (\bibinfo {year} {2015})},\ \Eprint {http://arxiv.org/abs/1507.04950} {arXiv:1507.04950 [hep-ph]} \BibitemShut {NoStop}%
\bibitem [{\citenamefont {Colangelo}\ \emph {et~al.}(2006)\citenamefont {Colangelo}, \citenamefont {Gasser}, \citenamefont {Kubis},\ and\ \citenamefont {Rusetsky}}]{Colangelo:2006va}%
  \BibitemOpen
  \bibfield  {author} {\bibinfo {author} {\bibfnamefont {Gilberto}\ \bibnamefont {Colangelo}}, \bibinfo {author} {\bibfnamefont {Juerg}\ \bibnamefont {Gasser}}, \bibinfo {author} {\bibfnamefont {Bastian}\ \bibnamefont {Kubis}}, \ and\ \bibinfo {author} {\bibfnamefont {Akaki}\ \bibnamefont {Rusetsky}},\ }\bibfield  {title} {\enquote {\bibinfo {title} {{Cusps in $K\to 3\pi$ decays}},}\ }\href {\doibase 10.1016/j.physletb.2006.05.017} {\bibfield  {journal} {\bibinfo  {journal} {Phys. Lett. B}\ }\textbf {\bibinfo {volume} {638}},\ \bibinfo {pages} {187--194} (\bibinfo {year} {2006})},\ \Eprint {http://arxiv.org/abs/hep-ph/0604084} {arXiv:hep-ph/0604084} \BibitemShut {NoStop}%
\bibitem [{\citenamefont {Gasser}\ \emph {et~al.}(2011)\citenamefont {Gasser}, \citenamefont {Kubis},\ and\ \citenamefont {Rusetsky}}]{Gasser:2011ju}%
  \BibitemOpen
  \bibfield  {author} {\bibinfo {author} {\bibfnamefont {Jurg}\ \bibnamefont {Gasser}}, \bibinfo {author} {\bibfnamefont {Bastian}\ \bibnamefont {Kubis}}, \ and\ \bibinfo {author} {\bibfnamefont {Akaki}\ \bibnamefont {Rusetsky}},\ }\bibfield  {title} {\enquote {\bibinfo {title} {{Cusps in $K\to 3\pi$ decays: a theoretical framework}},}\ }\href {\doibase 10.1016/j.nuclphysb.2011.04.013} {\bibfield  {journal} {\bibinfo  {journal} {Nucl. Phys. B}\ }\textbf {\bibinfo {volume} {850}},\ \bibinfo {pages} {96--147} (\bibinfo {year} {2011})},\ \Eprint {http://arxiv.org/abs/1103.4273} {arXiv:1103.4273 [hep-ph]} \BibitemShut {NoStop}%
\bibitem [{\citenamefont {Briscoe}\ \emph {et~al.}(2023)\citenamefont {Briscoe}, \citenamefont {Schmidt}, \citenamefont {Strakovsky}, \citenamefont {Workman},\ and\ \citenamefont {Svarc}}]{Briscoe:2023gmb}%
  \BibitemOpen
  \bibfield  {author} {\bibinfo {author} {\bibfnamefont {William~J.}\ \bibnamefont {Briscoe}}, \bibinfo {author} {\bibfnamefont {Axel}\ \bibnamefont {Schmidt}}, \bibinfo {author} {\bibfnamefont {Igor}\ \bibnamefont {Strakovsky}}, \bibinfo {author} {\bibfnamefont {Ron~L.}\ \bibnamefont {Workman}}, \ and\ \bibinfo {author} {\bibfnamefont {Alfred}\ \bibnamefont {Svarc}} (\bibinfo {collaboration} {SAID Group}),\ }\bibfield  {title} {\enquote {\bibinfo {title} {{Extended SAID partial-wave analysis of pion photoproduction}},}\ }\href {\doibase 10.1103/PhysRevC.108.065205} {\bibfield  {journal} {\bibinfo  {journal} {Phys. Rev. C}\ }\textbf {\bibinfo {volume} {108}},\ \bibinfo {pages} {065205} (\bibinfo {year} {2023})},\ \Eprint {http://arxiv.org/abs/2309.06631} {arXiv:2309.06631 [hep-ph]} \BibitemShut {NoStop}%
\bibitem [{\citenamefont {Aitchison}\ and\ \citenamefont {Kacser}(1964)}]{Aitchison:1964rwb}%
  \BibitemOpen
  \bibfield  {author} {\bibinfo {author} {\bibfnamefont {I.~J.~R.}\ \bibnamefont {Aitchison}}\ and\ \bibinfo {author} {\bibfnamefont {C.}~\bibnamefont {Kacser}},\ }\bibfield  {title} {\enquote {\bibinfo {title} {{Complex Propagators in Perturbation Theory}},}\ }\href {\doibase 10.1103/physrev.133.b1239} {\bibfield  {journal} {\bibinfo  {journal} {Phys. Rev.}\ }\textbf {\bibinfo {volume} {133}},\ \bibinfo {pages} {B1239--B1257} (\bibinfo {year} {1964})}\BibitemShut {NoStop}%
\bibitem [{\citenamefont {Ceci}\ \emph {et~al.}(2011)\citenamefont {Ceci}, \citenamefont {Doring}, \citenamefont {Hanhart}, \citenamefont {Krewald}, \citenamefont {Mei{\ss}ner},\ and\ \citenamefont {Svarc}}]{Ceci:2011ae}%
  \BibitemOpen
  \bibfield  {author} {\bibinfo {author} {\bibfnamefont {S.}~\bibnamefont {Ceci}}, \bibinfo {author} {\bibfnamefont {M.}~\bibnamefont {Doring}}, \bibinfo {author} {\bibfnamefont {C.}~\bibnamefont {Hanhart}}, \bibinfo {author} {\bibfnamefont {S.}~\bibnamefont {Krewald}}, \bibinfo {author} {\bibfnamefont {U.~G.}\ \bibnamefont {Mei{\ss}ner}}, \ and\ \bibinfo {author} {\bibfnamefont {A.}~\bibnamefont {Svarc}},\ }\bibfield  {title} {\enquote {\bibinfo {title} {{Relevance of complex branch points for partial wave analysis}},}\ }\href {\doibase 10.1103/PhysRevC.84.015205} {\bibfield  {journal} {\bibinfo  {journal} {Phys. Rev. C}\ }\textbf {\bibinfo {volume} {84}},\ \bibinfo {pages} {015205} (\bibinfo {year} {2011})},\ \Eprint {http://arxiv.org/abs/1104.3490} {arXiv:1104.3490 [nucl-th]} \BibitemShut {NoStop}%
\bibitem [{\citenamefont {Bayar}\ \emph {et~al.}(2016)\citenamefont {Bayar}, \citenamefont {Aceti}, \citenamefont {Guo},\ and\ \citenamefont {Oset}}]{Bayar:2016ftu}%
  \BibitemOpen
  \bibfield  {author} {\bibinfo {author} {\bibfnamefont {Melahat}\ \bibnamefont {Bayar}}, \bibinfo {author} {\bibfnamefont {Francesca}\ \bibnamefont {Aceti}}, \bibinfo {author} {\bibfnamefont {Feng-Kun}\ \bibnamefont {Guo}}, \ and\ \bibinfo {author} {\bibfnamefont {Eulogio}\ \bibnamefont {Oset}},\ }\bibfield  {title} {\enquote {\bibinfo {title} {{A Discussion on Triangle Singularities in the $\Lambda_b \to J/\psi K^{-} p$ Reaction}},}\ }\href {\doibase 10.1103/PhysRevD.94.074039} {\bibfield  {journal} {\bibinfo  {journal} {Phys. Rev. D}\ }\textbf {\bibinfo {volume} {94}},\ \bibinfo {pages} {074039} (\bibinfo {year} {2016})},\ \Eprint {http://arxiv.org/abs/1609.04133} {arXiv:1609.04133 [hep-ph]} \BibitemShut {NoStop}%
\bibitem [{\citenamefont {Guo}\ \emph {et~al.}(2020)\citenamefont {Guo}, \citenamefont {Liu},\ and\ \citenamefont {Sakai}}]{Guo:2019twa}%
  \BibitemOpen
  \bibfield  {author} {\bibinfo {author} {\bibfnamefont {Feng-Kun}\ \bibnamefont {Guo}}, \bibinfo {author} {\bibfnamefont {Xiao-Hai}\ \bibnamefont {Liu}}, \ and\ \bibinfo {author} {\bibfnamefont {Shuntaro}\ \bibnamefont {Sakai}},\ }\bibfield  {title} {\enquote {\bibinfo {title} {{Threshold cusps and triangle singularities in hadronic reactions}},}\ }\href {\doibase 10.1016/j.ppnp.2020.103757} {\bibfield  {journal} {\bibinfo  {journal} {Prog. Part. Nucl. Phys.}\ }\textbf {\bibinfo {volume} {112}},\ \bibinfo {pages} {103757} (\bibinfo {year} {2020})},\ \Eprint {http://arxiv.org/abs/1912.07030} {arXiv:1912.07030 [hep-ph]} \BibitemShut {NoStop}%
\bibitem [{\citenamefont {Isken}\ \emph {et~al.}(2024)\citenamefont {Isken}, \citenamefont {Guo}, \citenamefont {Heo}, \citenamefont {Korpa},\ and\ \citenamefont {Lutz}}]{Isken:2023xfo}%
  \BibitemOpen
  \bibfield  {author} {\bibinfo {author} {\bibfnamefont {Tobias}\ \bibnamefont {Isken}}, \bibinfo {author} {\bibfnamefont {Xiao-Yu}\ \bibnamefont {Guo}}, \bibinfo {author} {\bibfnamefont {Yonggoo}\ \bibnamefont {Heo}}, \bibinfo {author} {\bibfnamefont {Csaba~L.}\ \bibnamefont {Korpa}}, \ and\ \bibinfo {author} {\bibfnamefont {Matthias F.~M.}\ \bibnamefont {Lutz}},\ }\bibfield  {title} {\enquote {\bibinfo {title} {{Triangle and box diagrams in coupled-channel systems from the chiral Lagrangian}},}\ }\href {\doibase 10.1103/PhysRevD.109.034032} {\bibfield  {journal} {\bibinfo  {journal} {Phys. Rev. D}\ }\textbf {\bibinfo {volume} {109}},\ \bibinfo {pages} {034032} (\bibinfo {year} {2024})},\ \Eprint {http://arxiv.org/abs/2309.09695} {arXiv:2309.09695 [hep-ph]} \BibitemShut {NoStop}%
\bibitem [{\citenamefont {Debastiani}\ \emph {et~al.}(2019)\citenamefont {Debastiani}, \citenamefont {Sakai},\ and\ \citenamefont {Oset}}]{Debastiani:2018xoi}%
  \BibitemOpen
  \bibfield  {author} {\bibinfo {author} {\bibfnamefont {V.~R.}\ \bibnamefont {Debastiani}}, \bibinfo {author} {\bibfnamefont {S.}~\bibnamefont {Sakai}}, \ and\ \bibinfo {author} {\bibfnamefont {E.}~\bibnamefont {Oset}},\ }\bibfield  {title} {\enquote {\bibinfo {title} {{Considerations on the Schmid theorem for triangle singularities}},}\ }\href {\doibase 10.1140/epjc/s10052-019-6558-1} {\bibfield  {journal} {\bibinfo  {journal} {Eur. Phys. J. C}\ }\textbf {\bibinfo {volume} {79}},\ \bibinfo {pages} {69} (\bibinfo {year} {2019})},\ \Eprint {http://arxiv.org/abs/1809.06890} {arXiv:1809.06890 [hep-ph]} \BibitemShut {NoStop}%
\bibitem [{\citenamefont {Dai}\ \emph {et~al.}(2018)\citenamefont {Dai}, \citenamefont {Pavao}, \citenamefont {Sakai},\ and\ \citenamefont {Oset}}]{Dai:2018hqb}%
  \BibitemOpen
  \bibfield  {author} {\bibinfo {author} {\bibfnamefont {L.~R.}\ \bibnamefont {Dai}}, \bibinfo {author} {\bibfnamefont {R.}~\bibnamefont {Pavao}}, \bibinfo {author} {\bibfnamefont {S.}~\bibnamefont {Sakai}}, \ and\ \bibinfo {author} {\bibfnamefont {E.}~\bibnamefont {Oset}},\ }\bibfield  {title} {\enquote {\bibinfo {title} {{Anomalous enhancement of the isospin-violating $\Lambda(1405)$ production by a triangle singularity in $\Lambda_c\rightarrow\pi^+\pi^0\pi^0\Sigma^0$}},}\ }\href {\doibase 10.1103/PhysRevD.97.116004} {\bibfield  {journal} {\bibinfo  {journal} {Phys. Rev. D}\ }\textbf {\bibinfo {volume} {97}},\ \bibinfo {pages} {116004} (\bibinfo {year} {2018})},\ \Eprint {http://arxiv.org/abs/1804.01136} {arXiv:1804.01136 [hep-ph]} \BibitemShut {NoStop}%
\bibitem [{\citenamefont {Dai}\ \emph {et~al.}(2019)\citenamefont {Dai}, \citenamefont {Yu},\ and\ \citenamefont {Oset}}]{Dai:2018rra}%
  \BibitemOpen
  \bibfield  {author} {\bibinfo {author} {\bibfnamefont {L.~R.}\ \bibnamefont {Dai}}, \bibinfo {author} {\bibfnamefont {Q.~X.}\ \bibnamefont {Yu}}, \ and\ \bibinfo {author} {\bibfnamefont {E.}~\bibnamefont {Oset}},\ }\bibfield  {title} {\enquote {\bibinfo {title} {{Triangle singularity in $\tau^- \to \nu_\tau \pi^- f_0(980)$ ($a_0(980)$) decays}},}\ }\href {\doibase 10.1103/PhysRevD.99.016021} {\bibfield  {journal} {\bibinfo  {journal} {Phys. Rev. D}\ }\textbf {\bibinfo {volume} {99}},\ \bibinfo {pages} {016021} (\bibinfo {year} {2019})},\ \Eprint {http://arxiv.org/abs/1809.11007} {arXiv:1809.11007 [hep-ph]} \BibitemShut {NoStop}%
\bibitem [{\citenamefont {Liang}\ \emph {et~al.}(2019)\citenamefont {Liang}, \citenamefont {Chen}, \citenamefont {Oset},\ and\ \citenamefont {Wang}}]{Liang:2019jtr}%
  \BibitemOpen
  \bibfield  {author} {\bibinfo {author} {\bibfnamefont {Wei-Hong}\ \bibnamefont {Liang}}, \bibinfo {author} {\bibfnamefont {Hua-Xing}\ \bibnamefont {Chen}}, \bibinfo {author} {\bibfnamefont {Eulogio}\ \bibnamefont {Oset}}, \ and\ \bibinfo {author} {\bibfnamefont {En}~\bibnamefont {Wang}},\ }\bibfield  {title} {\enquote {\bibinfo {title} {{Triangle singularity in the $J/\psi \rightarrow K^+ K^- f_0(980)(a_0(980))$ decays}},}\ }\href {\doibase 10.1140/epjc/s10052-019-6928-8} {\bibfield  {journal} {\bibinfo  {journal} {Eur. Phys. J. C}\ }\textbf {\bibinfo {volume} {79}},\ \bibinfo {pages} {411} (\bibinfo {year} {2019})},\ \Eprint {http://arxiv.org/abs/1903.01252} {arXiv:1903.01252 [hep-ph]} \BibitemShut {NoStop}%
\bibitem [{\citenamefont {Jing}\ \emph {et~al.}(2019)\citenamefont {Jing}, \citenamefont {Sakai}, \citenamefont {Guo},\ and\ \citenamefont {Zou}}]{Jing:2019cbw}%
  \BibitemOpen
  \bibfield  {author} {\bibinfo {author} {\bibfnamefont {Hao-Jie}\ \bibnamefont {Jing}}, \bibinfo {author} {\bibfnamefont {Shuntaro}\ \bibnamefont {Sakai}}, \bibinfo {author} {\bibfnamefont {Feng-Kun}\ \bibnamefont {Guo}}, \ and\ \bibinfo {author} {\bibfnamefont {Bing-Song}\ \bibnamefont {Zou}},\ }\bibfield  {title} {\enquote {\bibinfo {title} {{Triangle singularities in ${J/\psi\rightarrow\eta\pi^0\phi}$ and ${\pi^0\pi^0\phi}$}},}\ }\href {\doibase 10.1103/PhysRevD.100.114010} {\bibfield  {journal} {\bibinfo  {journal} {Phys. Rev. D}\ }\textbf {\bibinfo {volume} {100}},\ \bibinfo {pages} {114010} (\bibinfo {year} {2019})},\ \Eprint {http://arxiv.org/abs/1907.12719} {arXiv:1907.12719 [hep-ph]} \BibitemShut {NoStop}%
\bibitem [{\citenamefont {Du}\ and\ \citenamefont {Zhao}(2021)}]{Du:2021zdg}%
  \BibitemOpen
  \bibfield  {author} {\bibinfo {author} {\bibfnamefont {Meng-Chuan}\ \bibnamefont {Du}}\ and\ \bibinfo {author} {\bibfnamefont {Qiang}\ \bibnamefont {Zhao}},\ }\bibfield  {title} {\enquote {\bibinfo {title} {{Comprehensive study of light axial vector mesons with the presence of triangle singularity}},}\ }\href {\doibase 10.1103/PhysRevD.104.036008} {\bibfield  {journal} {\bibinfo  {journal} {Phys. Rev. D}\ }\textbf {\bibinfo {volume} {104}},\ \bibinfo {pages} {036008} (\bibinfo {year} {2021})},\ \Eprint {http://arxiv.org/abs/2103.16861} {arXiv:2103.16861 [hep-ph]} \BibitemShut {NoStop}%
\bibitem [{\citenamefont {Duan}\ \emph {et~al.}(2024)\citenamefont {Duan}, \citenamefont {Qiu}, \citenamefont {Ling},\ and\ \citenamefont {Zhao}}]{Duan:2023dky}%
  \BibitemOpen
  \bibfield  {author} {\bibinfo {author} {\bibfnamefont {Ming-Xiao}\ \bibnamefont {Duan}}, \bibinfo {author} {\bibfnamefont {Lin}\ \bibnamefont {Qiu}}, \bibinfo {author} {\bibfnamefont {Xi-Zhe}\ \bibnamefont {Ling}}, \ and\ \bibinfo {author} {\bibfnamefont {Qiang}\ \bibnamefont {Zhao}},\ }\bibfield  {title} {\enquote {\bibinfo {title} {{Predictions for feed-down enhancements at the $\Lambda_cD$ and $\Lambda_cD^*$ thresholds via the triangle and box singularities}},}\ }\href {\doibase 10.1103/PhysRevD.109.L031507} {\bibfield  {journal} {\bibinfo  {journal} {Phys. Rev. D}\ }\textbf {\bibinfo {volume} {109}},\ \bibinfo {pages} {L031507} (\bibinfo {year} {2024})},\ \Eprint {http://arxiv.org/abs/2303.13329} {arXiv:2303.13329 [hep-ph]} \BibitemShut {NoStop}%
\bibitem [{\citenamefont {Wang}\ \emph {et~al.}(2017)\citenamefont {Wang}, \citenamefont {Xie}, \citenamefont {Liang}, \citenamefont {Guo},\ and\ \citenamefont {Oset}}]{Wang:2016dtb}%
  \BibitemOpen
  \bibfield  {author} {\bibinfo {author} {\bibfnamefont {En}~\bibnamefont {Wang}}, \bibinfo {author} {\bibfnamefont {Ju-Jun}\ \bibnamefont {Xie}}, \bibinfo {author} {\bibfnamefont {Wei-Hong}\ \bibnamefont {Liang}}, \bibinfo {author} {\bibfnamefont {Feng-Kun}\ \bibnamefont {Guo}}, \ and\ \bibinfo {author} {\bibfnamefont {Eulogio}\ \bibnamefont {Oset}},\ }\bibfield  {title} {\enquote {\bibinfo {title} {{Role of a triangle singularity in the $\gamma p\rightarrow K^+ \Lambda(1405)$ reaction}},}\ }\href {\doibase 10.1103/PhysRevC.95.015205} {\bibfield  {journal} {\bibinfo  {journal} {Phys. Rev. C}\ }\textbf {\bibinfo {volume} {95}},\ \bibinfo {pages} {015205} (\bibinfo {year} {2017})},\ \Eprint {http://arxiv.org/abs/1610.07117} {arXiv:1610.07117 [hep-ph]} \BibitemShut {NoStop}%
\bibitem [{\citenamefont {Nakamura}\ \emph {et~al.}(2023)\citenamefont {Nakamura}, \citenamefont {Li}, \citenamefont {Peng}, \citenamefont {Sun},\ and\ \citenamefont {Zhou}}]{Nakamura:2023obk}%
  \BibitemOpen
  \bibfield  {author} {\bibinfo {author} {\bibfnamefont {S.~X.}\ \bibnamefont {Nakamura}}, \bibinfo {author} {\bibfnamefont {X.~H.}\ \bibnamefont {Li}}, \bibinfo {author} {\bibfnamefont {H.~P.}\ \bibnamefont {Peng}}, \bibinfo {author} {\bibfnamefont {Z.~T.}\ \bibnamefont {Sun}}, \ and\ \bibinfo {author} {\bibfnamefont {X.~R.}\ \bibnamefont {Zhou}},\ }\bibfield  {title} {\enquote {\bibinfo {title} {{Global coupled-channel analysis of $e^+e^-\to c\bar{c}$ processes in $\sqrt{s}=3.75-4.7$ GeV}},}\ }\href@noop {} {\  (\bibinfo {year} {2023})},\ \Eprint {http://arxiv.org/abs/2312.17658} {arXiv:2312.17658 [hep-ph]} \BibitemShut {NoStop}%
\bibitem [{\citenamefont {Zhang}(2024)}]{Zhang:2024dth}%
  \BibitemOpen
  \bibfield  {author} {\bibinfo {author} {\bibfnamefont {Xu}~\bibnamefont {Zhang}},\ }\bibfield  {title} {\enquote {\bibinfo {title} {{Relativistic three-body scattering and the $D_0D^{*+}-D^+D^{*0}$ system}},}\ }\href {\doibase 10.1103/PhysRevD.109.094010} {\bibfield  {journal} {\bibinfo  {journal} {Phys. Rev. D}\ }\textbf {\bibinfo {volume} {109}},\ \bibinfo {pages} {094010} (\bibinfo {year} {2024})},\ \Eprint {http://arxiv.org/abs/2402.02151} {arXiv:2402.02151 [hep-ph]} \BibitemShut {NoStop}%
\bibitem [{\citenamefont {Achasov}\ and\ \citenamefont {Shestakov}(2022)}]{Achasov:2022onn}%
  \BibitemOpen
  \bibfield  {author} {\bibinfo {author} {\bibfnamefont {N.~N.}\ \bibnamefont {Achasov}}\ and\ \bibinfo {author} {\bibfnamefont {G.~N.}\ \bibnamefont {Shestakov}},\ }\bibfield  {title} {\enquote {\bibinfo {title} {{Triangle singularities in the $T_{cc}^+\rightarrow D^*+D_0 \rightarrow\pi+D_0D_0$ decay width}},}\ }\href {\doibase 10.1103/PhysRevD.105.096038} {\bibfield  {journal} {\bibinfo  {journal} {Phys. Rev. D}\ }\textbf {\bibinfo {volume} {105}},\ \bibinfo {pages} {096038} (\bibinfo {year} {2022})},\ \Eprint {http://arxiv.org/abs/2203.17100} {arXiv:2203.17100 [hep-ph]} \BibitemShut {NoStop}%
\bibitem [{\citenamefont {Nakamura}\ \emph {et~al.}(2024)\citenamefont {Nakamura}, \citenamefont {Huang}, \citenamefont {Wu}, \citenamefont {Peng}, \citenamefont {Zhang},\ and\ \citenamefont {Zhu}}]{Nakamura:2023hbt}%
  \BibitemOpen
  \bibfield  {author} {\bibinfo {author} {\bibfnamefont {S.~X.}\ \bibnamefont {Nakamura}}, \bibinfo {author} {\bibfnamefont {Q.}~\bibnamefont {Huang}}, \bibinfo {author} {\bibfnamefont {J.~J.}\ \bibnamefont {Wu}}, \bibinfo {author} {\bibfnamefont {H.~P.}\ \bibnamefont {Peng}}, \bibinfo {author} {\bibfnamefont {Y.}~\bibnamefont {Zhang}}, \ and\ \bibinfo {author} {\bibfnamefont {Y.~C.}\ \bibnamefont {Zhu}},\ }\bibfield  {title} {\enquote {\bibinfo {title} {{Three-body unitary coupled-channel approach to radiative $J/\psi$ decays and $\eta\,(1405/1475)$}},}\ }\href {\doibase 10.1103/PhysRevD.109.014021} {\bibfield  {journal} {\bibinfo  {journal} {Phys. Rev. D}\ }\textbf {\bibinfo {volume} {109}},\ \bibinfo {pages} {014021} (\bibinfo {year} {2024})},\ \Eprint {http://arxiv.org/abs/2311.05391} {arXiv:2311.05391 [hep-ph]} \BibitemShut {NoStop}%
\bibitem [{\citenamefont {Adolph}\ \emph {et~al.}(2015)\citenamefont {Adolph} \emph {et~al.}}]{COMPASS:2015kdx}%
  \BibitemOpen
  \bibfield  {author} {\bibinfo {author} {\bibfnamefont {C.}~\bibnamefont {Adolph}} \emph {et~al.} (\bibinfo {collaboration} {COMPASS}),\ }\bibfield  {title} {\enquote {\bibinfo {title} {{Observation of a New Narrow Axial-Vector Meson $a_1$(1420)}},}\ }\href {\doibase 10.1103/PhysRevLett.115.082001} {\bibfield  {journal} {\bibinfo  {journal} {Phys. Rev. Lett.}\ }\textbf {\bibinfo {volume} {115}},\ \bibinfo {pages} {082001} (\bibinfo {year} {2015})},\ \Eprint {http://arxiv.org/abs/1501.05732} {arXiv:1501.05732 [hep-ex]} \BibitemShut {NoStop}%
\bibitem [{\citenamefont {Rabusov}\ \emph {et~al.}(2023)\citenamefont {Rabusov}, \citenamefont {Greenwald},\ and\ \citenamefont {Paul}}]{Rabusov:2023tna}%
  \BibitemOpen
  \bibfield  {author} {\bibinfo {author} {\bibfnamefont {Andrei}\ \bibnamefont {Rabusov}}, \bibinfo {author} {\bibfnamefont {Daniel}\ \bibnamefont {Greenwald}}, \ and\ \bibinfo {author} {\bibfnamefont {Stephan}\ \bibnamefont {Paul}} (\bibinfo {collaboration} {Belle}),\ }\bibfield  {title} {\enquote {\bibinfo {title} {{Partial-wave analysis of $\tau^-\to\pi^-\pi^-\pi^+\nu_\tau$ at Belle}},}\ }in\ \href@noop {} {\emph {\bibinfo {booktitle} {{20th International Conference on Hadron Spectroscopy and Structure}}}}\ (\bibinfo {year} {2023})\ \Eprint {http://arxiv.org/abs/2310.09155} {arXiv:2310.09155 [hep-ex]} \BibitemShut {NoStop}%
\bibitem [{\citenamefont {Mikhasenko}\ \emph {et~al.}(2015)\citenamefont {Mikhasenko}, \citenamefont {Ketzer},\ and\ \citenamefont {Sarantsev}}]{Mikhasenko:2015oxp}%
  \BibitemOpen
  \bibfield  {author} {\bibinfo {author} {\bibfnamefont {M.}~\bibnamefont {Mikhasenko}}, \bibinfo {author} {\bibfnamefont {B.}~\bibnamefont {Ketzer}}, \ and\ \bibinfo {author} {\bibfnamefont {Andrey}\ \bibnamefont {Sarantsev}},\ }\bibfield  {title} {\enquote {\bibinfo {title} {{Nature of the $a_1(1420)$}},}\ }\href {\doibase 10.1103/PhysRevD.91.094015} {\bibfield  {journal} {\bibinfo  {journal} {Phys. Rev. D}\ }\textbf {\bibinfo {volume} {91}},\ \bibinfo {pages} {094015} (\bibinfo {year} {2015})},\ \Eprint {http://arxiv.org/abs/1501.07023} {arXiv:1501.07023 [hep-ph]} \BibitemShut {NoStop}%
\bibitem [{\citenamefont {Aceti}\ \emph {et~al.}(2016)\citenamefont {Aceti}, \citenamefont {Dai},\ and\ \citenamefont {Oset}}]{Aceti:2016yeb}%
  \BibitemOpen
  \bibfield  {author} {\bibinfo {author} {\bibfnamefont {F.}~\bibnamefont {Aceti}}, \bibinfo {author} {\bibfnamefont {L.~R.}\ \bibnamefont {Dai}}, \ and\ \bibinfo {author} {\bibfnamefont {E.}~\bibnamefont {Oset}},\ }\bibfield  {title} {\enquote {\bibinfo {title} {{$a_1(1420)$ peak as the $\pi f_0(980)$ decay mode of the $a_1(1260)$}},}\ }\href {\doibase 10.1103/PhysRevD.94.096015} {\bibfield  {journal} {\bibinfo  {journal} {Phys. Rev. D}\ }\textbf {\bibinfo {volume} {94}},\ \bibinfo {pages} {096015} (\bibinfo {year} {2016})},\ \Eprint {http://arxiv.org/abs/1606.06893} {arXiv:1606.06893 [hep-ph]} \BibitemShut {NoStop}%
\bibitem [{\citenamefont {Alexeev}\ \emph {et~al.}(2021)\citenamefont {Alexeev} \emph {et~al.}}]{COMPASS:2020yhb}%
  \BibitemOpen
  \bibfield  {author} {\bibinfo {author} {\bibfnamefont {G.~D.}\ \bibnamefont {Alexeev}} \emph {et~al.} (\bibinfo {collaboration} {COMPASS}),\ }\bibfield  {title} {\enquote {\bibinfo {title} {{Triangle Singularity as the Origin of the $a_1(1420)$}},}\ }\href {\doibase 10.1103/PhysRevLett.127.082501} {\bibfield  {journal} {\bibinfo  {journal} {Phys. Rev. Lett.}\ }\textbf {\bibinfo {volume} {127}},\ \bibinfo {pages} {082501} (\bibinfo {year} {2021})},\ \Eprint {http://arxiv.org/abs/2006.05342} {arXiv:2006.05342 [hep-ph]} \BibitemShut {NoStop}%
\bibitem [{\citenamefont {Mai}\ \emph {et~al.}(2017)\citenamefont {Mai}, \citenamefont {Hu}, \citenamefont {Doring}, \citenamefont {Pilloni},\ and\ \citenamefont {Szczepaniak}}]{Mai:2017vot}%
  \BibitemOpen
  \bibfield  {author} {\bibinfo {author} {\bibfnamefont {M.}~\bibnamefont {Mai}}, \bibinfo {author} {\bibfnamefont {B.}~\bibnamefont {Hu}}, \bibinfo {author} {\bibfnamefont {M.}~\bibnamefont {Doring}}, \bibinfo {author} {\bibfnamefont {A.}~\bibnamefont {Pilloni}}, \ and\ \bibinfo {author} {\bibfnamefont {A.}~\bibnamefont {Szczepaniak}},\ }\bibfield  {title} {\enquote {\bibinfo {title} {{Three-body Unitarity with Isobars Revisited}},}\ }\href {\doibase 10.1140/epja/i2017-12368-4} {\bibfield  {journal} {\bibinfo  {journal} {Eur. Phys. J. A}\ }\textbf {\bibinfo {volume} {53}},\ \bibinfo {pages} {177} (\bibinfo {year} {2017})},\ \Eprint {http://arxiv.org/abs/1706.06118} {arXiv:1706.06118 [nucl-th]} \BibitemShut {NoStop}%
\bibitem [{\citenamefont {Sadasivan}\ \emph {et~al.}(2022)\citenamefont {Sadasivan}, \citenamefont {Alexandru}, \citenamefont {Akdag}, \citenamefont {Amorim}, \citenamefont {Brett}, \citenamefont {Culver}, \citenamefont {D\"oring}, \citenamefont {Lee},\ and\ \citenamefont {Mai}}]{Sadasivan:2021emk}%
  \BibitemOpen
  \bibfield  {author} {\bibinfo {author} {\bibfnamefont {Daniel}\ \bibnamefont {Sadasivan}}, \bibinfo {author} {\bibfnamefont {Andrei}\ \bibnamefont {Alexandru}}, \bibinfo {author} {\bibfnamefont {Hakan}\ \bibnamefont {Akdag}}, \bibinfo {author} {\bibfnamefont {Felipe}\ \bibnamefont {Amorim}}, \bibinfo {author} {\bibfnamefont {Ruair\'\i{}}\ \bibnamefont {Brett}}, \bibinfo {author} {\bibfnamefont {Chris}\ \bibnamefont {Culver}}, \bibinfo {author} {\bibfnamefont {Michael}\ \bibnamefont {D\"oring}}, \bibinfo {author} {\bibfnamefont {Frank~X.}\ \bibnamefont {Lee}}, \ and\ \bibinfo {author} {\bibfnamefont {Maxim}\ \bibnamefont {Mai}},\ }\bibfield  {title} {\enquote {\bibinfo {title} {{Pole position of the $a_1(1260)$ resonance in a three-body unitary framework}},}\ }\href {\doibase 10.1103/PhysRevD.105.054020} {\bibfield  {journal} {\bibinfo  {journal} {Phys. Rev. D}\ }\textbf {\bibinfo {volume} {105}},\ \bibinfo {pages} {054020} (\bibinfo {year} {2022})},\ \Eprint {http://arxiv.org/abs/2112.03355}
  {arXiv:2112.03355 [hep-ph]} \BibitemShut {NoStop}%
\bibitem [{\citenamefont {Faddeev}(1960)}]{Faddeev:1960su}%
  \BibitemOpen
  \bibfield  {author} {\bibinfo {author} {\bibfnamefont {L.~D.}\ \bibnamefont {Faddeev}},\ }\bibfield  {title} {\enquote {\bibinfo {title} {{Scattering theory for a three particle system}},}\ }\href@noop {} {\bibfield  {journal} {\bibinfo  {journal} {Zh. Eksp. Teor. Fiz.}\ }\textbf {\bibinfo {volume} {39}},\ \bibinfo {pages} {1459--1467} (\bibinfo {year} {1960})}\BibitemShut {NoStop}%
\bibitem [{\citenamefont {Mai}\ and\ \citenamefont {D\"oring}(2017)}]{Mai:2017bge}%
  \BibitemOpen
  \bibfield  {author} {\bibinfo {author} {\bibfnamefont {M.}~\bibnamefont {Mai}}\ and\ \bibinfo {author} {\bibfnamefont {M.}~\bibnamefont {D\"oring}},\ }\bibfield  {title} {\enquote {\bibinfo {title} {{Three-body Unitarity in the Finite Volume}},}\ }\href {\doibase 10.1140/epja/i2017-12440-1} {\bibfield  {journal} {\bibinfo  {journal} {Eur. Phys. J. A}\ }\textbf {\bibinfo {volume} {53}},\ \bibinfo {pages} {240} (\bibinfo {year} {2017})},\ \Eprint {http://arxiv.org/abs/1709.08222} {arXiv:1709.08222 [hep-lat]} \BibitemShut {NoStop}%
\bibitem [{\citenamefont {Mai}\ \emph {et~al.}(2021{\natexlab{a}})\citenamefont {Mai}, \citenamefont {D\"oring},\ and\ \citenamefont {Rusetsky}}]{Mai:2021lwb}%
  \BibitemOpen
  \bibfield  {author} {\bibinfo {author} {\bibfnamefont {Maxim}\ \bibnamefont {Mai}}, \bibinfo {author} {\bibfnamefont {Michael}\ \bibnamefont {D\"oring}}, \ and\ \bibinfo {author} {\bibfnamefont {Akaki}\ \bibnamefont {Rusetsky}},\ }\bibfield  {title} {\enquote {\bibinfo {title} {{Multi-particle systems on the lattice and chiral extrapolations: a brief review}},}\ }\href {\doibase 10.1140/epjs/s11734-021-00146-5} {\bibfield  {journal} {\bibinfo  {journal} {Eur. Phys. J. ST}\ }\textbf {\bibinfo {volume} {230}},\ \bibinfo {pages} {1623--1643} (\bibinfo {year} {2021}{\natexlab{a}})},\ \Eprint {http://arxiv.org/abs/2103.00577} {arXiv:2103.00577 [hep-lat]} \BibitemShut {NoStop}%
\bibitem [{\citenamefont {Mai}\ \emph {et~al.}(2021{\natexlab{b}})\citenamefont {Mai}, \citenamefont {Alexandru}, \citenamefont {Brett}, \citenamefont {Culver}, \citenamefont {D\"oring}, \citenamefont {Lee},\ and\ \citenamefont {Sadasivan}}]{Mai:2021nul}%
  \BibitemOpen
  \bibfield  {author} {\bibinfo {author} {\bibfnamefont {Maxim}\ \bibnamefont {Mai}}, \bibinfo {author} {\bibfnamefont {Andrei}\ \bibnamefont {Alexandru}}, \bibinfo {author} {\bibfnamefont {Ruair\'\i{}}\ \bibnamefont {Brett}}, \bibinfo {author} {\bibfnamefont {Chris}\ \bibnamefont {Culver}}, \bibinfo {author} {\bibfnamefont {Michael}\ \bibnamefont {D\"oring}}, \bibinfo {author} {\bibfnamefont {Frank~X.}\ \bibnamefont {Lee}}, \ and\ \bibinfo {author} {\bibfnamefont {Daniel}\ \bibnamefont {Sadasivan}} (\bibinfo {collaboration} {GWQCD}),\ }\bibfield  {title} {\enquote {\bibinfo {title} {{Three-Body Dynamics of the $a_1(1260)$ Resonance from Lattice QCD}},}\ }\href {\doibase 10.1103/PhysRevLett.127.222001} {\bibfield  {journal} {\bibinfo  {journal} {Phys. Rev. Lett.}\ }\textbf {\bibinfo {volume} {127}},\ \bibinfo {pages} {222001} (\bibinfo {year} {2021}{\natexlab{b}})},\ \Eprint {http://arxiv.org/abs/2107.03973} {arXiv:2107.03973 [hep-lat]} \BibitemShut {NoStop}%
\bibitem [{\citenamefont {Yan}\ \emph {et~al.}(2024)\citenamefont {Yan}, \citenamefont {Garofalo}, \citenamefont {Mai}, \citenamefont {Mei\ss{}ner},\ and\ \citenamefont {Urbach}}]{Yan:2024gwp}%
  \BibitemOpen
  \bibfield  {author} {\bibinfo {author} {\bibfnamefont {Haobo}\ \bibnamefont {Yan}}, \bibinfo {author} {\bibfnamefont {Marco}\ \bibnamefont {Garofalo}}, \bibinfo {author} {\bibfnamefont {Maxim}\ \bibnamefont {Mai}}, \bibinfo {author} {\bibfnamefont {Ulf-G.}\ \bibnamefont {Mei\ss{}ner}}, \ and\ \bibinfo {author} {\bibfnamefont {Carsten}\ \bibnamefont {Urbach}},\ }\bibfield  {title} {\enquote {\bibinfo {title} {{The $\omega$-meson from lattice QCD}},}\ }\href@noop {} {\  (\bibinfo {year} {2024})},\ \Eprint {http://arxiv.org/abs/2407.16659} {arXiv:2407.16659 [hep-lat]} \BibitemShut {NoStop}%
\bibitem [{\citenamefont {Doring}\ \emph {et~al.}(2009)\citenamefont {Doring}, \citenamefont {Hanhart}, \citenamefont {Huang}, \citenamefont {Krewald},\ and\ \citenamefont {Mei{\ss}ner}}]{Doring:2009yv}%
  \BibitemOpen
  \bibfield  {author} {\bibinfo {author} {\bibfnamefont {M.}~\bibnamefont {Doring}}, \bibinfo {author} {\bibfnamefont {C.}~\bibnamefont {Hanhart}}, \bibinfo {author} {\bibfnamefont {F.}~\bibnamefont {Huang}}, \bibinfo {author} {\bibfnamefont {S.}~\bibnamefont {Krewald}}, \ and\ \bibinfo {author} {\bibfnamefont {U.~G.}\ \bibnamefont {Mei{\ss}ner}},\ }\bibfield  {title} {\enquote {\bibinfo {title} {{Analytic properties of the scattering amplitude and resonances parameters in a meson exchange model}},}\ }\href {\doibase 10.1016/j.nuclphysa.2009.08.010} {\bibfield  {journal} {\bibinfo  {journal} {Nucl. Phys. A}\ }\textbf {\bibinfo {volume} {829}},\ \bibinfo {pages} {170--209} (\bibinfo {year} {2009})},\ \Eprint {http://arxiv.org/abs/0903.4337} {arXiv:0903.4337 [nucl-th]} \BibitemShut {NoStop}%
\bibitem [{\citenamefont {Suzuki}\ \emph {et~al.}(2010)\citenamefont {Suzuki}, \citenamefont {Julia-Diaz}, \citenamefont {Kamano}, \citenamefont {Lee}, \citenamefont {Matsuyama},\ and\ \citenamefont {Sato}}]{Suzuki:2009nj}%
  \BibitemOpen
  \bibfield  {author} {\bibinfo {author} {\bibfnamefont {N.}~\bibnamefont {Suzuki}}, \bibinfo {author} {\bibfnamefont {B.}~\bibnamefont {Julia-Diaz}}, \bibinfo {author} {\bibfnamefont {H.}~\bibnamefont {Kamano}}, \bibinfo {author} {\bibfnamefont {T.~S.~H.}\ \bibnamefont {Lee}}, \bibinfo {author} {\bibfnamefont {A.}~\bibnamefont {Matsuyama}}, \ and\ \bibinfo {author} {\bibfnamefont {T.}~\bibnamefont {Sato}},\ }\bibfield  {title} {\enquote {\bibinfo {title} {{Disentangling the Dynamical Origin of $P_{11}$ Nucleon Resonances}},}\ }\href {\doibase 10.1103/PhysRevLett.104.042302} {\bibfield  {journal} {\bibinfo  {journal} {Phys. Rev. Lett.}\ }\textbf {\bibinfo {volume} {104}},\ \bibinfo {pages} {042302} (\bibinfo {year} {2010})},\ \Eprint {http://arxiv.org/abs/0909.1356} {arXiv:0909.1356 [nucl-th]} \BibitemShut {NoStop}%
\bibitem [{\citenamefont {Jackura}\ \emph {et~al.}(2021)\citenamefont {Jackura}, \citenamefont {Brice\~no}, \citenamefont {Dawid}, \citenamefont {Islam},\ and\ \citenamefont {McCarty}}]{Jackura:2020bsk}%
  \BibitemOpen
  \bibfield  {author} {\bibinfo {author} {\bibfnamefont {Andrew~W.}\ \bibnamefont {Jackura}}, \bibinfo {author} {\bibfnamefont {Ra\'ul~A.}\ \bibnamefont {Brice\~no}}, \bibinfo {author} {\bibfnamefont {Sebastian~M.}\ \bibnamefont {Dawid}}, \bibinfo {author} {\bibfnamefont {Md~Habib~E.}\ \bibnamefont {Islam}}, \ and\ \bibinfo {author} {\bibfnamefont {Connor}\ \bibnamefont {McCarty}},\ }\bibfield  {title} {\enquote {\bibinfo {title} {{Solving relativistic three-body integral equations in the presence of bound states}},}\ }\href {\doibase 10.1103/PhysRevD.104.014507} {\bibfield  {journal} {\bibinfo  {journal} {Phys. Rev. D}\ }\textbf {\bibinfo {volume} {104}},\ \bibinfo {pages} {014507} (\bibinfo {year} {2021})},\ \Eprint {http://arxiv.org/abs/2010.09820} {arXiv:2010.09820 [hep-lat]} \BibitemShut {NoStop}%
\bibitem [{\citenamefont {Dawid}\ \emph {et~al.}(2023)\citenamefont {Dawid}, \citenamefont {Islam},\ and\ \citenamefont {Brice\~no}}]{Dawid:2023jrj}%
  \BibitemOpen
  \bibfield  {author} {\bibinfo {author} {\bibfnamefont {Sebastian~M.}\ \bibnamefont {Dawid}}, \bibinfo {author} {\bibfnamefont {Md~Habib~E.}\ \bibnamefont {Islam}}, \ and\ \bibinfo {author} {\bibfnamefont {Ra\'ul~A.}\ \bibnamefont {Brice\~no}},\ }\bibfield  {title} {\enquote {\bibinfo {title} {{Analytic continuation of the relativistic three-particle scattering amplitudes}},}\ }\href {\doibase 10.1103/PhysRevD.108.034016} {\bibfield  {journal} {\bibinfo  {journal} {Phys. Rev. D}\ }\textbf {\bibinfo {volume} {108}},\ \bibinfo {pages} {034016} (\bibinfo {year} {2023})},\ \Eprint {http://arxiv.org/abs/2303.04394} {arXiv:2303.04394 [nucl-th]} \BibitemShut {NoStop}%
\bibitem [{\citenamefont {Dawid}\ \emph {et~al.}(2024)\citenamefont {Dawid}, \citenamefont {Islam}, \citenamefont {Briceno},\ and\ \citenamefont {Jackura}}]{Dawid:2023kxu}%
  \BibitemOpen
  \bibfield  {author} {\bibinfo {author} {\bibfnamefont {Sebastian~M.}\ \bibnamefont {Dawid}}, \bibinfo {author} {\bibfnamefont {Md~Habib~E.}\ \bibnamefont {Islam}}, \bibinfo {author} {\bibfnamefont {Raul~A.}\ \bibnamefont {Briceno}}, \ and\ \bibinfo {author} {\bibfnamefont {Andrew~W.}\ \bibnamefont {Jackura}},\ }\bibfield  {title} {\enquote {\bibinfo {title} {{Evolution of Efimov states}},}\ }\href {\doibase 10.1103/PhysRevA.109.043325} {\bibfield  {journal} {\bibinfo  {journal} {Phys. Rev. A}\ }\textbf {\bibinfo {volume} {109}},\ \bibinfo {pages} {043325} (\bibinfo {year} {2024})},\ \Eprint {http://arxiv.org/abs/2309.01732} {arXiv:2309.01732 [nucl-th]} \BibitemShut {NoStop}%
\bibitem [{\citenamefont {Garofalo}\ \emph {et~al.}(2023)\citenamefont {Garofalo}, \citenamefont {Mai}, \citenamefont {Romero-L\'opez}, \citenamefont {Rusetsky},\ and\ \citenamefont {Urbach}}]{Garofalo:2022pux}%
  \BibitemOpen
  \bibfield  {author} {\bibinfo {author} {\bibfnamefont {Marco}\ \bibnamefont {Garofalo}}, \bibinfo {author} {\bibfnamefont {Maxim}\ \bibnamefont {Mai}}, \bibinfo {author} {\bibfnamefont {Fernando}\ \bibnamefont {Romero-L\'opez}}, \bibinfo {author} {\bibfnamefont {Akaki}\ \bibnamefont {Rusetsky}}, \ and\ \bibinfo {author} {\bibfnamefont {Carsten}\ \bibnamefont {Urbach}},\ }\bibfield  {title} {\enquote {\bibinfo {title} {{Three-body resonances in the $\varphi^4$ theory}},}\ }\href {\doibase 10.1007/JHEP02(2023)252} {\bibfield  {journal} {\bibinfo  {journal} {JHEP}\ }\textbf {\bibinfo {volume} {02}},\ \bibinfo {pages} {252} (\bibinfo {year} {2023})},\ \Eprint {http://arxiv.org/abs/2211.05605} {arXiv:2211.05605 [hep-lat]} \BibitemShut {NoStop}%
\bibitem [{\citenamefont {Stamen}\ \emph {et~al.}(2023)\citenamefont {Stamen}, \citenamefont {Isken}, \citenamefont {Kubis}, \citenamefont {Mikhasenko},\ and\ \citenamefont {Niehus}}]{Stamen:2022eda}%
  \BibitemOpen
  \bibfield  {author} {\bibinfo {author} {\bibfnamefont {Dominik}\ \bibnamefont {Stamen}}, \bibinfo {author} {\bibfnamefont {Tobias}\ \bibnamefont {Isken}}, \bibinfo {author} {\bibfnamefont {Bastian}\ \bibnamefont {Kubis}}, \bibinfo {author} {\bibfnamefont {Mikhail}\ \bibnamefont {Mikhasenko}}, \ and\ \bibinfo {author} {\bibfnamefont {Malwin}\ \bibnamefont {Niehus}},\ }\bibfield  {title} {\enquote {\bibinfo {title} {{Analysis of rescattering effects in $3\pi $ final states}},}\ }\href {\doibase 10.1140/epjc/s10052-023-11665-x} {\bibfield  {journal} {\bibinfo  {journal} {Eur. Phys. J. C}\ }\textbf {\bibinfo {volume} {83}},\ \bibinfo {pages} {510} (\bibinfo {year} {2023})},\ \bibinfo {note} {[Erratum: Eur.Phys.J.C 83, 586 (2023)]},\ \Eprint {http://arxiv.org/abs/2212.11767} {arXiv:2212.11767 [hep-ph]} \BibitemShut {NoStop}%
\bibitem [{\citenamefont {Lucha}\ \emph {et~al.}(2007)\citenamefont {Lucha}, \citenamefont {Melikhov},\ and\ \citenamefont {Simula}}]{Lucha:2006vc}%
  \BibitemOpen
  \bibfield  {author} {\bibinfo {author} {\bibfnamefont {Wolfgang}\ \bibnamefont {Lucha}}, \bibinfo {author} {\bibfnamefont {Dmitri}\ \bibnamefont {Melikhov}}, \ and\ \bibinfo {author} {\bibfnamefont {Silvano}\ \bibnamefont {Simula}},\ }\bibfield  {title} {\enquote {\bibinfo {title} {{Dispersion representations and anomalous singularities of the triangle diagram}},}\ }\href {\doibase 10.1103/PhysRevD.75.016001} {\bibfield  {journal} {\bibinfo  {journal} {Phys. Rev. D}\ }\textbf {\bibinfo {volume} {75}},\ \bibinfo {pages} {016001} (\bibinfo {year} {2007})},\ \bibinfo {note} {[Erratum: Phys.Rev.D 92, 019901 (2015)]},\ \Eprint {http://arxiv.org/abs/hep-ph/0610330} {arXiv:hep-ph/0610330} \BibitemShut {NoStop}%
\bibitem [{\citenamefont {{Y. Feng, F. Gil, A. S. Sakthivasan, et al.}}(2025)}]{Fengprep}%
  \BibitemOpen
  \bibfield  {author} {\bibinfo {author} {\bibnamefont {{Y. Feng, F. Gil, A. S. Sakthivasan, et al.}}},\ }\href@noop {} {\enquote {\bibinfo {title} {{In Preparation}},}\ } (\bibinfo {year} {2025})\BibitemShut {NoStop}%
\bibitem [{\citenamefont {Workman}\ \emph {et~al.}(2022)\citenamefont {Workman} \emph {et~al.}}]{ParticleDataGroup:2022pth}%
  \BibitemOpen
  \bibfield  {author} {\bibinfo {author} {\bibfnamefont {R.~L.}\ \bibnamefont {Workman}} \emph {et~al.} (\bibinfo {collaboration} {Particle Data Group}),\ }\bibfield  {title} {\enquote {\bibinfo {title} {{Review of Particle Physics}},}\ }\href {\doibase 10.1093/ptep/ptac097} {\bibfield  {journal} {\bibinfo  {journal} {PTEP}\ }\textbf {\bibinfo {volume} {2022}},\ \bibinfo {pages} {083C01} (\bibinfo {year} {2022})}\BibitemShut {NoStop}%
\bibitem [{\citenamefont {Liu}\ \emph {et~al.}(2024)\citenamefont {Liu}, \citenamefont {Pan}, \citenamefont {Liu}, \citenamefont {Wu}, \citenamefont {Lu},\ and\ \citenamefont {Geng}}]{Liu:2024uxn}%
  \BibitemOpen
  \bibfield  {author} {\bibinfo {author} {\bibfnamefont {Ming-Zhu}\ \bibnamefont {Liu}}, \bibinfo {author} {\bibfnamefont {Ya-Wen}\ \bibnamefont {Pan}}, \bibinfo {author} {\bibfnamefont {Zhi-Wei}\ \bibnamefont {Liu}}, \bibinfo {author} {\bibfnamefont {Tian-Wei}\ \bibnamefont {Wu}}, \bibinfo {author} {\bibfnamefont {Jun-Xu}\ \bibnamefont {Lu}}, \ and\ \bibinfo {author} {\bibfnamefont {Li-Sheng}\ \bibnamefont {Geng}},\ }\bibfield  {title} {\enquote {\bibinfo {title} {{Three ways to decipher the nature of exotic hadrons: multiplets, three-body hadronic molecules, and correlation functions}},}\ }\href@noop {} {\  (\bibinfo {year} {2024})},\ \Eprint {http://arxiv.org/abs/2404.06399} {arXiv:2404.06399 [hep-ph]} \BibitemShut {NoStop}%
\bibitem [{\citenamefont {Albaladejo}\ \emph {et~al.}(2022)\citenamefont {Albaladejo} \emph {et~al.}}]{JPAC:2021rxu}%
  \BibitemOpen
  \bibfield  {author} {\bibinfo {author} {\bibfnamefont {Miguel}\ \bibnamefont {Albaladejo}} \emph {et~al.} (\bibinfo {collaboration} {JPAC}),\ }\bibfield  {title} {\enquote {\bibinfo {title} {{Novel approaches in hadron spectroscopy}},}\ }\href {\doibase 10.1016/j.ppnp.2022.103981} {\bibfield  {journal} {\bibinfo  {journal} {Prog. Part. Nucl. Phys.}\ }\textbf {\bibinfo {volume} {127}},\ \bibinfo {pages} {103981} (\bibinfo {year} {2022})},\ \Eprint {http://arxiv.org/abs/2112.13436} {arXiv:2112.13436 [hep-ph]} \BibitemShut {NoStop}%
\bibitem [{\citenamefont {Ketzer}\ \emph {et~al.}(2020)\citenamefont {Ketzer}, \citenamefont {Grube},\ and\ \citenamefont {Ryabchikov}}]{Ketzer:2019wmd}%
  \BibitemOpen
  \bibfield  {author} {\bibinfo {author} {\bibfnamefont {Bernhard}\ \bibnamefont {Ketzer}}, \bibinfo {author} {\bibfnamefont {Boris}\ \bibnamefont {Grube}}, \ and\ \bibinfo {author} {\bibfnamefont {Dmitry}\ \bibnamefont {Ryabchikov}},\ }\bibfield  {title} {\enquote {\bibinfo {title} {{Light-Meson Spectroscopy with COMPASS}},}\ }\href {\doibase 10.1016/j.ppnp.2020.103755} {\bibfield  {journal} {\bibinfo  {journal} {Prog. Part. Nucl. Phys.}\ }\textbf {\bibinfo {volume} {113}},\ \bibinfo {pages} {103755} (\bibinfo {year} {2020})},\ \Eprint {http://arxiv.org/abs/1909.06366} {arXiv:1909.06366 [hep-ex]} \BibitemShut {NoStop}%
\bibitem [{\citenamefont {Aghasyan}\ \emph {et~al.}(2018)\citenamefont {Aghasyan} \emph {et~al.}}]{COMPASS:2018uzl}%
  \BibitemOpen
  \bibfield  {author} {\bibinfo {author} {\bibfnamefont {M.}~\bibnamefont {Aghasyan}} \emph {et~al.} (\bibinfo {collaboration} {COMPASS}),\ }\bibfield  {title} {\enquote {\bibinfo {title} {{Light isovector resonances in $\pi^- p \to \pi^-\pi^-\pi^+ p$ at 190 GeV/$c$}},}\ }\href {\doibase 10.1103/PhysRevD.98.092003} {\bibfield  {journal} {\bibinfo  {journal} {Phys. Rev. D}\ }\textbf {\bibinfo {volume} {98}},\ \bibinfo {pages} {092003} (\bibinfo {year} {2018})},\ \Eprint {http://arxiv.org/abs/1802.05913} {arXiv:1802.05913 [hep-ex]} \BibitemShut {NoStop}%
\bibitem [{\citenamefont {Albrecht}\ \emph {et~al.}(2020)\citenamefont {Albrecht} \emph {et~al.}}]{CrystalBarrel:2019zqh}%
  \BibitemOpen
  \bibfield  {author} {\bibinfo {author} {\bibfnamefont {M.}~\bibnamefont {Albrecht}} \emph {et~al.} (\bibinfo {collaboration} {Crystal Barrel}),\ }\bibfield  {title} {\enquote {\bibinfo {title} {{Coupled channel analysis of ${\bar{p}p}\,\rightarrow \,\pi ^0\pi ^0\eta $, ${\pi ^0\eta \eta }$ and ${K^+K^-\pi ^0}$ at 900 MeV/c and of ${\pi \pi }$-scattering data}},}\ }\href {\doibase 10.1140/epjc/s10052-020-7930-x} {\bibfield  {journal} {\bibinfo  {journal} {Eur. Phys. J. C}\ }\textbf {\bibinfo {volume} {80}},\ \bibinfo {pages} {453} (\bibinfo {year} {2020})},\ \Eprint {http://arxiv.org/abs/1909.07091} {arXiv:1909.07091 [hep-ex]} \BibitemShut {NoStop}%
\bibitem [{\citenamefont {Ablikim}\ \emph {et~al.}(2023)\citenamefont {Ablikim} \emph {et~al.}}]{BESIII:2023qgj}%
  \BibitemOpen
  \bibfield  {author} {\bibinfo {author} {\bibfnamefont {Medina}\ \bibnamefont {Ablikim}} \emph {et~al.} (\bibinfo {collaboration} {BESIII}),\ }\bibfield  {title} {\enquote {\bibinfo {title} {{Amplitude analysis and branching fraction measurement of the decay $D^+\to K_S^0\pi^+\pi^0\pi^0$}},}\ }\href {\doibase 10.1007/JHEP09(2023)077} {\bibfield  {journal} {\bibinfo  {journal} {JHEP}\ }\textbf {\bibinfo {volume} {09}},\ \bibinfo {pages} {077} (\bibinfo {year} {2023})},\ \Eprint {http://arxiv.org/abs/2305.15879} {arXiv:2305.15879 [hep-ex]} \BibitemShut {NoStop}%
\bibitem [{\citenamefont {Rabusov}\ \emph {et~al.}(2022)\citenamefont {Rabusov}, \citenamefont {Greenwald},\ and\ \citenamefont {Paul}}]{Rabusov:2022woa}%
  \BibitemOpen
  \bibfield  {author} {\bibinfo {author} {\bibfnamefont {Andrei}\ \bibnamefont {Rabusov}}, \bibinfo {author} {\bibfnamefont {Daniel}\ \bibnamefont {Greenwald}}, \ and\ \bibinfo {author} {\bibfnamefont {Stephan}\ \bibnamefont {Paul}},\ }\bibfield  {title} {\enquote {\bibinfo {title} {{Partial wave analysis of $\tau^-\to\pi^-\pi^+\pi^-\nu_\tau$ at Belle}},}\ }\href {\doibase 10.22323/1.414.1034} {\bibfield  {journal} {\bibinfo  {journal} {PoS}\ }\textbf {\bibinfo {volume} {ICHEP2022}},\ \bibinfo {pages} {1034} (\bibinfo {year} {2022})},\ \Eprint {http://arxiv.org/abs/2211.11696} {arXiv:2211.11696 [hep-ex]} \BibitemShut {NoStop}%
\bibitem [{\citenamefont {Aaij}\ \emph {et~al.}(2023)\citenamefont {Aaij} \emph {et~al.}}]{LHCb:2022lja}%
  \BibitemOpen
  \bibfield  {author} {\bibinfo {author} {\bibfnamefont {R.}~\bibnamefont {Aaij}} \emph {et~al.} (\bibinfo {collaboration} {LHCb}),\ }\bibfield  {title} {\enquote {\bibinfo {title} {{Amplitude analysis of the $D^+\to\pi^- \pi^+\pi^+$ decay and measurement of the $\pi^-\pi^+$ S-wave amplitude}},}\ }\href {\doibase 10.1007/JHEP06(2023)044} {\bibfield  {journal} {\bibinfo  {journal} {JHEP}\ }\textbf {\bibinfo {volume} {06}},\ \bibinfo {pages} {044} (\bibinfo {year} {2023})},\ \Eprint {http://arxiv.org/abs/2208.03300} {arXiv:2208.03300 [hep-ex]} \BibitemShut {NoStop}%
\bibitem [{\citenamefont {Landau}(1959)}]{Landau:1959fi}%
  \BibitemOpen
  \bibfield  {author} {\bibinfo {author} {\bibfnamefont {L.~D.}\ \bibnamefont {Landau}},\ }\bibfield  {title} {\enquote {\bibinfo {title} {{On analytic properties of vertex parts in quantum field theory}},}\ }\href {\doibase 10.1016/B978-0-08-010586-4.50103-6} {\bibfield  {journal} {\bibinfo  {journal} {Nucl. Phys.}\ }\textbf {\bibinfo {volume} {13}},\ \bibinfo {pages} {181--192} (\bibinfo {year} {1959})}\BibitemShut {NoStop}%
\bibitem [{\citenamefont {Polkinghorne}\ and\ \citenamefont {Screaton}(1960{\natexlab{a}})}]{Polkinghorne:1960cjb}%
  \BibitemOpen
  \bibfield  {author} {\bibinfo {author} {\bibfnamefont {J.~C.}\ \bibnamefont {Polkinghorne}}\ and\ \bibinfo {author} {\bibfnamefont {G.~R.}\ \bibnamefont {Screaton}},\ }\bibfield  {title} {\enquote {\bibinfo {title} {{The analytic properties of perturbation theory \textemdash{} I}},}\ }\href {\doibase 10.1007/bf02860252} {\bibfield  {journal} {\bibinfo  {journal} {Nuovo Cim.}\ }\textbf {\bibinfo {volume} {15}},\ \bibinfo {pages} {289--300} (\bibinfo {year} {1960}{\natexlab{a}})}\BibitemShut {NoStop}%
\bibitem [{\citenamefont {Polkinghorne}\ and\ \citenamefont {Screaton}(1960{\natexlab{b}})}]{Polkinghorne:1960udx}%
  \BibitemOpen
  \bibfield  {author} {\bibinfo {author} {\bibfnamefont {J.~C.}\ \bibnamefont {Polkinghorne}}\ and\ \bibinfo {author} {\bibfnamefont {G.~R.}\ \bibnamefont {Screaton}},\ }\bibfield  {title} {\enquote {\bibinfo {title} {{The analytic properties of perturbation theory \textemdash{} II}},}\ }\href {\doibase 10.1007/bf02860197} {\bibfield  {journal} {\bibinfo  {journal} {Nuovo Cim.}\ }\textbf {\bibinfo {volume} {15}},\ \bibinfo {pages} {925--931} (\bibinfo {year} {1960}{\natexlab{b}})}\BibitemShut {NoStop}%
\bibitem [{\citenamefont {Coleman}\ and\ \citenamefont {Norton}(1965)}]{Coleman:1965xm}%
  \BibitemOpen
  \bibfield  {author} {\bibinfo {author} {\bibfnamefont {S.}~\bibnamefont {Coleman}}\ and\ \bibinfo {author} {\bibfnamefont {R.~E.}\ \bibnamefont {Norton}},\ }\bibfield  {title} {\enquote {\bibinfo {title} {{Singularities in the physical region}},}\ }\href {\doibase 10.1007/BF02750472} {\bibfield  {journal} {\bibinfo  {journal} {Nuovo Cim.}\ }\textbf {\bibinfo {volume} {38}},\ \bibinfo {pages} {438--442} (\bibinfo {year} {1965})}\BibitemShut {NoStop}%
\bibitem [{\citenamefont {Nauenberg}\ and\ \citenamefont {Pais}(1962)}]{PhysRev.126.360}%
  \BibitemOpen
  \bibfield  {author} {\bibinfo {author} {\bibfnamefont {M.}~\bibnamefont {Nauenberg}}\ and\ \bibinfo {author} {\bibfnamefont {A.}~\bibnamefont {Pais}},\ }\bibfield  {title} {\enquote {\bibinfo {title} {Woolly cusps},}\ }\href {\doibase 10.1103/PhysRev.126.360} {\bibfield  {journal} {\bibinfo  {journal} {Phys. Rev.}\ }\textbf {\bibinfo {volume} {126}},\ \bibinfo {pages} {360--364} (\bibinfo {year} {1962})}\BibitemShut {NoStop}%
\bibitem [{\citenamefont {Sadasivan}\ \emph {et~al.}(2020)\citenamefont {Sadasivan}, \citenamefont {Mai}, \citenamefont {Akdag},\ and\ \citenamefont {D\"oring}}]{Sadasivan:2020syi}%
  \BibitemOpen
  \bibfield  {author} {\bibinfo {author} {\bibfnamefont {D.}~\bibnamefont {Sadasivan}}, \bibinfo {author} {\bibfnamefont {M.}~\bibnamefont {Mai}}, \bibinfo {author} {\bibfnamefont {H.}~\bibnamefont {Akdag}}, \ and\ \bibinfo {author} {\bibfnamefont {M.}~\bibnamefont {D\"oring}},\ }\bibfield  {title} {\enquote {\bibinfo {title} {{Dalitz plots and lineshape of $a_1(1260)$ from a relativistic three-body unitary approach}},}\ }\href {\doibase 10.1103/PhysRevD.101.094018} {\bibfield  {journal} {\bibinfo  {journal} {Phys. Rev. D}\ }\textbf {\bibinfo {volume} {101}},\ \bibinfo {pages} {094018} (\bibinfo {year} {2020})},\ \bibinfo {note} {[Erratum: Phys.Rev.D 103, 019901 (2021)]},\ \Eprint {http://arxiv.org/abs/2002.12431} {arXiv:2002.12431 [nucl-th]} \BibitemShut {NoStop}%
\bibitem [{\citenamefont {Brett}\ \emph {et~al.}(2021)\citenamefont {Brett}, \citenamefont {Culver}, \citenamefont {Mai}, \citenamefont {Alexandru}, \citenamefont {D\"oring},\ and\ \citenamefont {Lee}}]{Brett:2021wyd}%
  \BibitemOpen
  \bibfield  {author} {\bibinfo {author} {\bibfnamefont {Ruair\'\i{}}\ \bibnamefont {Brett}}, \bibinfo {author} {\bibfnamefont {Chris}\ \bibnamefont {Culver}}, \bibinfo {author} {\bibfnamefont {Maxim}\ \bibnamefont {Mai}}, \bibinfo {author} {\bibfnamefont {Andrei}\ \bibnamefont {Alexandru}}, \bibinfo {author} {\bibfnamefont {Michael}\ \bibnamefont {D\"oring}}, \ and\ \bibinfo {author} {\bibfnamefont {Frank~X.}\ \bibnamefont {Lee}},\ }\bibfield  {title} {\enquote {\bibinfo {title} {{Three-body interactions from the finite-volume QCD spectrum}},}\ }\href {\doibase 10.1103/PhysRevD.104.014501} {\bibfield  {journal} {\bibinfo  {journal} {Phys. Rev. D}\ }\textbf {\bibinfo {volume} {104}},\ \bibinfo {pages} {014501} (\bibinfo {year} {2021})},\ \Eprint {http://arxiv.org/abs/2101.06144} {arXiv:2101.06144 [hep-lat]} \BibitemShut {NoStop}%
\bibitem [{\citenamefont {Culver}\ \emph {et~al.}(2020)\citenamefont {Culver}, \citenamefont {Mai}, \citenamefont {Brett}, \citenamefont {Alexandru},\ and\ \citenamefont {D\"oring}}]{Culver:2019vvu}%
  \BibitemOpen
  \bibfield  {author} {\bibinfo {author} {\bibfnamefont {Chris}\ \bibnamefont {Culver}}, \bibinfo {author} {\bibfnamefont {Maxim}\ \bibnamefont {Mai}}, \bibinfo {author} {\bibfnamefont {Ruair\'\i{}}\ \bibnamefont {Brett}}, \bibinfo {author} {\bibfnamefont {Andrei}\ \bibnamefont {Alexandru}}, \ and\ \bibinfo {author} {\bibfnamefont {Michael}\ \bibnamefont {D\"oring}},\ }\bibfield  {title} {\enquote {\bibinfo {title} {{Three pion spectrum in the $I=3$ channel from lattice QCD}},}\ }\href {\doibase 10.1103/PhysRevD.101.114507} {\bibfield  {journal} {\bibinfo  {journal} {Phys. Rev. D}\ }\textbf {\bibinfo {volume} {101}},\ \bibinfo {pages} {114507} (\bibinfo {year} {2020})},\ \Eprint {http://arxiv.org/abs/1911.09047} {arXiv:1911.09047 [hep-lat]} \BibitemShut {NoStop}%
\bibitem [{\citenamefont {Alexandru}\ \emph {et~al.}(2020)\citenamefont {Alexandru}, \citenamefont {Brett}, \citenamefont {Culver}, \citenamefont {D\"oring}, \citenamefont {Guo}, \citenamefont {Lee},\ and\ \citenamefont {Mai}}]{Alexandru:2020xqf}%
  \BibitemOpen
  \bibfield  {author} {\bibinfo {author} {\bibfnamefont {Andrei}\ \bibnamefont {Alexandru}}, \bibinfo {author} {\bibfnamefont {Ruair\'\i{}}\ \bibnamefont {Brett}}, \bibinfo {author} {\bibfnamefont {Chris}\ \bibnamefont {Culver}}, \bibinfo {author} {\bibfnamefont {Michael}\ \bibnamefont {D\"oring}}, \bibinfo {author} {\bibfnamefont {Dehua}\ \bibnamefont {Guo}}, \bibinfo {author} {\bibfnamefont {Frank~X.}\ \bibnamefont {Lee}}, \ and\ \bibinfo {author} {\bibfnamefont {Maxim}\ \bibnamefont {Mai}},\ }\bibfield  {title} {\enquote {\bibinfo {title} {{Finite-volume energy spectrum of the $K^-K^-K^-$ system}},}\ }\href {\doibase 10.1103/PhysRevD.102.114523} {\bibfield  {journal} {\bibinfo  {journal} {Phys. Rev. D}\ }\textbf {\bibinfo {volume} {102}},\ \bibinfo {pages} {114523} (\bibinfo {year} {2020})},\ \Eprint {http://arxiv.org/abs/2009.12358} {arXiv:2009.12358 [hep-lat]} \BibitemShut {NoStop}%
\bibitem [{\citenamefont {Mai}\ and\ \citenamefont {Doring}(2019)}]{Mai:2018djl}%
  \BibitemOpen
  \bibfield  {author} {\bibinfo {author} {\bibfnamefont {Maxim}\ \bibnamefont {Mai}}\ and\ \bibinfo {author} {\bibfnamefont {Michael}\ \bibnamefont {Doring}},\ }\bibfield  {title} {\enquote {\bibinfo {title} {{Finite-Volume Spectrum of $\pi^+\pi^+$ and $\pi^+\pi^+\pi^+$ Systems}},}\ }\href {\doibase 10.1103/PhysRevLett.122.062503} {\bibfield  {journal} {\bibinfo  {journal} {Phys. Rev. Lett.}\ }\textbf {\bibinfo {volume} {122}},\ \bibinfo {pages} {062503} (\bibinfo {year} {2019})},\ \Eprint {http://arxiv.org/abs/1807.04746} {arXiv:1807.04746 [hep-lat]} \BibitemShut {NoStop}%
\bibitem [{\citenamefont {Feng}\ \emph {et~al.}(2024)\citenamefont {Feng}, \citenamefont {Gil}, \citenamefont {D\"oring}, \citenamefont {Molina}, \citenamefont {Mai}, \citenamefont {Shastry},\ and\ \citenamefont {Szczepaniak}}]{Feng:2024wyg}%
  \BibitemOpen
  \bibfield  {author} {\bibinfo {author} {\bibfnamefont {Yuchuan}\ \bibnamefont {Feng}}, \bibinfo {author} {\bibfnamefont {Fernando}\ \bibnamefont {Gil}}, \bibinfo {author} {\bibfnamefont {Michael}\ \bibnamefont {D\"oring}}, \bibinfo {author} {\bibfnamefont {Raquel}\ \bibnamefont {Molina}}, \bibinfo {author} {\bibfnamefont {Maxim}\ \bibnamefont {Mai}}, \bibinfo {author} {\bibfnamefont {Vanamali}\ \bibnamefont {Shastry}}, \ and\ \bibinfo {author} {\bibfnamefont {Adam}\ \bibnamefont {Szczepaniak}},\ }\bibfield  {title} {\enquote {\bibinfo {title} {{A unitary coupled-channel three-body amplitude with pions and kaons}},}\ }\href@noop {} {\  (\bibinfo {year} {2024})},\ \Eprint {http://arxiv.org/abs/2407.08721} {arXiv:2407.08721 [nucl-th]} \BibitemShut {NoStop}%
\bibitem [{\citenamefont {D\"oring}\ \emph {et~al.}(2018)\citenamefont {D\"oring}, \citenamefont {Hammer}, \citenamefont {Mai}, \citenamefont {Pang}, \citenamefont {Rusetsky},\ and\ \citenamefont {Wu}}]{Doring:2018xxx}%
  \BibitemOpen
  \bibfield  {author} {\bibinfo {author} {\bibfnamefont {M.}~\bibnamefont {D\"oring}}, \bibinfo {author} {\bibfnamefont {H.~W.}\ \bibnamefont {Hammer}}, \bibinfo {author} {\bibfnamefont {M.}~\bibnamefont {Mai}}, \bibinfo {author} {\bibfnamefont {J.~Y.}\ \bibnamefont {Pang}}, \bibinfo {author} {\bibfnamefont {A.}~\bibnamefont {Rusetsky}}, \ and\ \bibinfo {author} {\bibfnamefont {J.}~\bibnamefont {Wu}},\ }\bibfield  {title} {\enquote {\bibinfo {title} {{Three-body spectrum in a finite volume: the role of cubic symmetry}},}\ }\href {\doibase 10.1103/PhysRevD.97.114508} {\bibfield  {journal} {\bibinfo  {journal} {Phys. Rev. D}\ }\textbf {\bibinfo {volume} {97}},\ \bibinfo {pages} {114508} (\bibinfo {year} {2018})},\ \Eprint {http://arxiv.org/abs/1802.03362} {arXiv:1802.03362 [hep-lat]} \BibitemShut {NoStop}%
\bibitem [{\citenamefont {Ablikim}\ \emph {et~al.}(2005)\citenamefont {Ablikim} \emph {et~al.}}]{BES:2004twe}%
  \BibitemOpen
  \bibfield  {author} {\bibinfo {author} {\bibfnamefont {M.}~\bibnamefont {Ablikim}} \emph {et~al.} (\bibinfo {collaboration} {BES}),\ }\bibfield  {title} {\enquote {\bibinfo {title} {{Resonances in $J /\psi \to \phi \pi^+ \pi^-$ and $\phi K^+ K^-$}},}\ }\href {\doibase 10.1016/j.physletb.2004.12.041} {\bibfield  {journal} {\bibinfo  {journal} {Phys. Lett. B}\ }\textbf {\bibinfo {volume} {607}},\ \bibinfo {pages} {243--253} (\bibinfo {year} {2005})},\ \Eprint {http://arxiv.org/abs/hep-ex/0411001} {arXiv:hep-ex/0411001} \BibitemShut {NoStop}%
\bibitem [{\citenamefont {Aubert}\ \emph {et~al.}(2006)\citenamefont {Aubert} \emph {et~al.}}]{BaBar:2006hyf}%
  \BibitemOpen
  \bibfield  {author} {\bibinfo {author} {\bibfnamefont {Bernard}\ \bibnamefont {Aubert}} \emph {et~al.} (\bibinfo {collaboration} {BaBar}),\ }\bibfield  {title} {\enquote {\bibinfo {title} {{Dalitz plot analysis of the decay $B^\pm \to K^\pm K^\pm K^\mp$}},}\ }\href {\doibase 10.1103/PhysRevD.74.032003} {\bibfield  {journal} {\bibinfo  {journal} {Phys. Rev. D}\ }\textbf {\bibinfo {volume} {74}},\ \bibinfo {pages} {032003} (\bibinfo {year} {2006})},\ \Eprint {http://arxiv.org/abs/hep-ex/0605003} {arXiv:hep-ex/0605003} \BibitemShut {NoStop}%
\bibitem [{\citenamefont {Hetherington}\ and\ \citenamefont {Schick}(1965)}]{Hetherington:1965zza}%
  \BibitemOpen
  \bibfield  {author} {\bibinfo {author} {\bibfnamefont {J.~H.}\ \bibnamefont {Hetherington}}\ and\ \bibinfo {author} {\bibfnamefont {L.~H.}\ \bibnamefont {Schick}},\ }\bibfield  {title} {\enquote {\bibinfo {title} {{Exact Multiple-Scattering Analysis of Low-Energy Elastic $K-d$ Scattering with Separable Potentials}},}\ }\href {\doibase 10.1103/PhysRev.137.B935} {\bibfield  {journal} {\bibinfo  {journal} {Phys. Rev.}\ }\textbf {\bibinfo {volume} {137}},\ \bibinfo {pages} {B935--B948} (\bibinfo {year} {1965})}\BibitemShut {NoStop}%
\bibitem [{\citenamefont {Cahill}\ and\ \citenamefont {Sloan}(1971)}]{Cahill:1971ddy}%
  \BibitemOpen
  \bibfield  {author} {\bibinfo {author} {\bibfnamefont {R.~T.}\ \bibnamefont {Cahill}}\ and\ \bibinfo {author} {\bibfnamefont {I.~H.}\ \bibnamefont {Sloan}},\ }\bibfield  {title} {\enquote {\bibinfo {title} {{Theory of neutron-deuteron break-up at 14.4 MeV}},}\ }\href {\doibase 10.1016/0375-9474(72)90798-1} {\bibfield  {journal} {\bibinfo  {journal} {Nucl. Phys. A}\ }\textbf {\bibinfo {volume} {165}},\ \bibinfo {pages} {161--179} (\bibinfo {year} {1971})},\ \bibinfo {note} {[Erratum: Nucl.Phys.A 196, 632--632 (1972)]}\BibitemShut {NoStop}%
\bibitem [{\citenamefont {Schmid}\ and\ \citenamefont {Ziegelmann}(1974)}]{SchmidZiegelmann}%
  \BibitemOpen
  \bibfield  {author} {\bibinfo {author} {\bibfnamefont {Erich~W.}\ \bibnamefont {Schmid}}\ and\ \bibinfo {author} {\bibfnamefont {Horst}\ \bibnamefont {Ziegelmann}},\ }\href@noop {} {\emph {\bibinfo {title} {The Quantum Mechanical Three-Body Problem}}}\ (\bibinfo  {publisher} {Friedr. Vieweg \& Sohn},\ \bibinfo {address} {Braunschweig},\ \bibinfo {year} {1974})\BibitemShut {NoStop}%
\bibitem [{\citenamefont {Adhikari}\ and\ \citenamefont {Amado}(1974)}]{Adhikari:1974fh}%
  \BibitemOpen
  \bibfield  {author} {\bibinfo {author} {\bibfnamefont {S.~K.}\ \bibnamefont {Adhikari}}\ and\ \bibinfo {author} {\bibfnamefont {R.~D.}\ \bibnamefont {Amado}},\ }\bibfield  {title} {\enquote {\bibinfo {title} {{Singularities in three-body final state amplitudes}},}\ }\href {\doibase 10.1103/PhysRevD.9.1467} {\bibfield  {journal} {\bibinfo  {journal} {Phys. Rev. D}\ }\textbf {\bibinfo {volume} {9}},\ \bibinfo {pages} {1467--1475} (\bibinfo {year} {1974})}\BibitemShut {NoStop}%
\bibitem [{\citenamefont {Matsuyama}\ \emph {et~al.}(2007)\citenamefont {Matsuyama}, \citenamefont {Sato},\ and\ \citenamefont {Lee}}]{Matsuyama:2006rp}%
  \BibitemOpen
  \bibfield  {author} {\bibinfo {author} {\bibfnamefont {A.}~\bibnamefont {Matsuyama}}, \bibinfo {author} {\bibfnamefont {T.}~\bibnamefont {Sato}}, \ and\ \bibinfo {author} {\bibfnamefont {T.~S.~H.}\ \bibnamefont {Lee}},\ }\bibfield  {title} {\enquote {\bibinfo {title} {{Dynamical coupled-channel model of meson production reactions in the nucleon resonance region}},}\ }\href {\doibase 10.1016/j.physrep.2006.12.003} {\bibfield  {journal} {\bibinfo  {journal} {Phys. Rept.}\ }\textbf {\bibinfo {volume} {439}},\ \bibinfo {pages} {193--253} (\bibinfo {year} {2007})},\ \Eprint {http://arxiv.org/abs/nucl-th/0608051} {arXiv:nucl-th/0608051} \BibitemShut {NoStop}%
\bibitem [{\citenamefont {Angell}(2010)}]{ANGELL2010904}%
  \BibitemOpen
  \bibfield  {author} {\bibinfo {author} {\bibfnamefont {David}\ \bibnamefont {Angell}},\ }\bibfield  {title} {\enquote {\bibinfo {title} {A family of continued fractions},}\ }\href {\doibase https://doi.org/10.1016/j.jnt.2009.12.003} {\bibfield  {journal} {\bibinfo  {journal} {Journal of Number Theory}\ }\textbf {\bibinfo {volume} {130}},\ \bibinfo {pages} {904--911} (\bibinfo {year} {2010})}\BibitemShut {NoStop}%
\bibitem [{\citenamefont {Lorentzen}(2010)}]{LORENTZEN20101364}%
  \BibitemOpen
  \bibfield  {author} {\bibinfo {author} {\bibfnamefont {Lisa}\ \bibnamefont {Lorentzen}},\ }\bibfield  {title} {\enquote {\bibinfo {title} {Padé approximation and continued fractions},}\ }\href {\doibase https://doi.org/10.1016/j.apnum.2010.03.016} {\bibfield  {journal} {\bibinfo  {journal} {Applied Numerical Mathematics}\ }\textbf {\bibinfo {volume} {60}},\ \bibinfo {pages} {1364--1370} (\bibinfo {year} {2010})},\ \bibinfo {note} {approximation and extrapolation of convergent and divergent sequences and series (CIRM, Luminy - France, 2009)}\BibitemShut {NoStop}%
\bibitem [{\citenamefont {{Abramowitz, M., and Stegun, I.}}(1964)}]{Abramowitz-Stegun}%
  \BibitemOpen
  \bibfield  {author} {\bibinfo {author} {\bibnamefont {{Abramowitz, M., and Stegun, I.}}},\ }\href@noop {} {\emph {\bibinfo {title} {{Handbook of Mathematical Functions with Formulas, Graphs, and Mathematical Tables}}}}\ (\bibinfo  {publisher} {Dover},\ \bibinfo {address} {New York},\ \bibinfo {year} {1964})\BibitemShut {NoStop}%
\bibitem [{\citenamefont {Pang}\ \emph {et~al.}(2024)\citenamefont {Pang}, \citenamefont {Bubna}, \citenamefont {M\"uller}, \citenamefont {Rusetsky},\ and\ \citenamefont {Wu}}]{Pang:2023jri}%
  \BibitemOpen
  \bibfield  {author} {\bibinfo {author} {\bibfnamefont {Jin-Yi}\ \bibnamefont {Pang}}, \bibinfo {author} {\bibfnamefont {Rishabh}\ \bibnamefont {Bubna}}, \bibinfo {author} {\bibfnamefont {Fabian}\ \bibnamefont {M\"uller}}, \bibinfo {author} {\bibfnamefont {Akaki}\ \bibnamefont {Rusetsky}}, \ and\ \bibinfo {author} {\bibfnamefont {Jia-Jun}\ \bibnamefont {Wu}},\ }\bibfield  {title} {\enquote {\bibinfo {title} {{Lellouch-L\"uscher factor for the $K\to3\pi$ decays}},}\ }\href {\doibase 10.1007/JHEP05(2024)269} {\bibfield  {journal} {\bibinfo  {journal} {JHEP}\ }\textbf {\bibinfo {volume} {05}},\ \bibinfo {pages} {269} (\bibinfo {year} {2024})},\ \Eprint {http://arxiv.org/abs/2312.04391} {arXiv:2312.04391 [hep-lat]} \BibitemShut {NoStop}%
\end{thebibliography}%
\clearpage
\appendix
\end{document}